\documentclass[a4paper,fleqn,usenatbib]{mnras}
\pdfoutput = 1

\usepackage[T1]{fontenc}
\usepackage{ae,aecompl}

\usepackage{graphicx}	
\graphicspath{{./Figures/}} 
\usepackage{grffile} 
\usepackage{amsmath}	
\usepackage{amssymb}	
\usepackage{bm} 
\usepackage{subfig} 
\usepackage{txfonts}
\usepackage{hyperref} 

\newcommand{\Nbg}{\ensuremath{N_\mathrm{bg}}} 
\newcommand{\Nc}{\ensuremath{N_\mathrm{c}}} 
\newcommand{\Ntot}{\ensuremath{N_\mathrm{total}}} 
\newcommand{\s}{\ensuremath{\bm{s}}} 
\newcommand{\A}{\ensuremath{\bm{A}}} 

\newenvironment{br}{}{}  
\newenvironment{br2}{}{}  

\title[Spatial Statistics in Star Forming Regions]{Spatial Statistics in Star Forming Regions:\\ Testing the Limits of Randomness}

\author[B. Retter et al.]{
B. Retter,
J. Hatchell
and Tim Naylor
\\
Physics and Astronomy, University of Exeter, Stocker Road, Exeter, EX4 4QL, UK}

\date{Accepted XXX. Received YYY; in original form ZZZ}

\pubyear{2019}

\begin{document}
\label{firstpage}
\pagerange{\pageref{firstpage}--\pageref{lastpage}}
\maketitle

\begin{abstract}
Observational studies of star formation reveal spatial distributions of Young Stellar Objects (YSOs) that are `snapshots' of an ongoing star formation process.  Using methods from spatial statistics it is possible to test the likelihood that a given distribution process could produce the observed patterns of YSOs. The aim of this paper is to determine the usefulness of the spatial statistics tests Diggle's G function (G), the `free-space' function (F), Ripley's K and O-ring for application to astrophysical data.  The spatial statistics tests were applied to simulated data containing 2D Gaussian clusters projected on random distributions of stars. The number of stars within the Gaussian cluster and number of background stars were varied to determine the tests' ability to reject complete spatial randomness (CSR) with changing signal-to-noise. \begin{br}The best performing test was O-ring optimised with overlapping logarithmic bins, closely followed by Ripley's K. The O-ring test is equivalent to the 2-point correlation function.  Both F and G (and the minimum spanning tree, \begin{br2}of which G is a subset)\end{br2} performed significantly less well, requiring a cluster with a factor of two higher signal-to-noise in order to reject CSR consistently.  We demonstrate the tests on example astrophysical datasets drawn from the \textit{Spitzer} catalogue.\end{br}
\end{abstract}




\section{INTRODUCTION}
\label{sec:Intro}
Star formation occurs within molecular clouds, and using the estimated ages and locations of young stellar objects (YSOs) it is possible to attempt a description of how the distribution of YSOs within a cloud may have evolved. In general, earlier stage YSOs are more densely clustered and situated closer to the densest regions of a cloud, while the more evolved YSOs tend to be more dispersed and further removed from dense gas \citep{polychroni2013}. \begin{br}However, the process producing this difference is uncertain. Possible explanations include migration of YSOs from their site of formation \citep{mairs2016,covey2006}, with older YSOs having had more time to move away, or star formation locations that change with time as the molecular clouds reconfigure themselves \citep{ybarra2013}.\end{br}

\begin{br}Studies of star formation frequently involve investigating clusters. These clusters are often identified by eye with cluster members then selected by an algorithm such as the minimum spanning tree (MST) which is the unique set of edges connecting a collection of points that minimises the total edge length with no cycles \citep{Gutermuth2009}; or the "friends-of-friends" algorithm which defines groups by collecting all members with separations no longer than a pre-defined length \citep{More2011,Huchra1982}. Such methods will identify clusters without consideration as to whether the points have been aggregated due to a physical process or if the cluster is purely coincidental, an overdensity of randomly distributed stars. Using spatial statistics it is possible to identify clustering within a dataset as well as determine if the clustering is statistically significant. \end{br}

\begin{br}Spatial statistics provides methods for testing the suitability of a model for distributing YSOs within a region. These models, known as spatial point processes within spatial statistics, are stochastic mechanisms for positioning points within a study widow informed by knowledge of the system in question. The first spatial point process that data are typically checked against is that of complete spatial randomness (CSR), where the location of any individual star\footnote{Within spatial statistics the locations of objects such as stars, galaxies or YSOs are referred to as events; for readability in this paper they will be referred to as stars.} 
is entirely random. This is because CSR represents a state of complete non-interaction between stars, and between stars and their environment and is a good baseline for proceeding to more complex processes.\end{br}

Within astrophysics, statistics such as the two point correlation function (2PCF) and MST tend to be used to classify the degree of clustering within a system rather than testing if the distribution has been produced by random processes. This is not surprising, as systems within astrophysics are not usually assumed to be entirely random. 
On the other hand, methods from spatial statistics are applied frequently in ecology and epidemiology, to determine if locations of trees or disease occurrences could be correlated \citep{Barot1999,Wiegand2009,Velazquez2016}.

This paper will focus on four of the most commonly applied methods; Diggle's G function (G), the `free-space function' (F), Ripley's K function (K) and the O-ring statistic (O). The aim is to investigate the ability of these statistical methods, discussed further in Section \ref{sec:Methods}, to reject CSR when the dataset contains both potential cluster members as well as randomly-distributed background objects. The results of these trials as well as the application to astrophysical data are discussed in Section \ref{sec:Results}. The tests are then compared to each other, the 2PCF and MST in Section \ref{sec:discussion}.

\section{THE STATISTICAL TESTS}
\label{sec:Methods}

When examining spatial point patterns, one method of analysis is to calculate a summary statistic for the pattern and determine whether the measured value is consistent with some null hypothesis. As discussed in Section \ref{sec:Intro}, often the first null hypothesis is that of CSR, a pattern which can be produced with a homogeneous Poisson point process. For CSR the distribution of the number of stars in regions $\s$ within the study space follows a Poisson distribution with mean $\lambda |\s|$, where $\lambda$ is the first-order intensity of the process and $|\s|$ is the area contained within $\s$. \begin{br}The first-order intensity is a measure of the number of stars per unit area, which for a stationary pattern such as CSR is constant across the entire study space.\end{br}

\subsection{First-Order Statistics}
First-order effects produce large-scale variation in the positioning of stars, making the \begin{br}first-order intensity\end{br} a function of position, $\lambda (x)|\s|$. These effects are typically environmental as the probability of a star being in region $\s$ has no dependence on neighbouring stars. For CSR the probability of a star having a nearest neighbour distance less than or equal to $w$ is $1-e^{-\pi \lambda w^2}$, which can be tested by looking at the \begin{br}nearest-neighbour distribution of the data estimated using Diggle's G function and the free-space function F.\end{br}

The first test, Diggle's G function, is an estimate of the cumulative probability distribution of nearest neighbour distances between stars. For a given distance, $\mathit{w}$, the uncorrected G($\mathit{w}$) is the number of stars with a nearest neighbour closer than $\mathit{w}$ divided by the total number of stars \citep{Diggle2013}; see Fig. \ref{fig:stat_explain}. 
\begin{br}However, an edge correction is needed to compensate for stars closer to the boundaries of the test region having fewer neighbours. \begin{br2}We have chosen to use the border method of edge-correction due to its intuitive nature and effectiveness. The border method functions by using stars with a distance to the closest boundary, $\mathit{b_i}$, greater than $\mathit{w}$ to estimate G($\mathit{w}$) while allowing stars within the border to be counted as a nearest neighbour \citep{Dale2014}.\end{br2} The region where $b_i \leq w$ is shown in Fig. \ref{fig:stat_explain} by a green border. While only rectangular-shaped windows are used in this paper the border-correction method is valid for arbitrarily shaped borders and is one of many edge-correction methods used for G and F, including the Kaplan-Meier estimator \citep{baddeley1997}.\end{br}

The free-space function (F) is similar to G except it is an estimate of the cumulative probability distribution of nearest neighbour distances between randomised positions in the study window and their nearest star, which will be referred to as $\mathit{x_i}$ (see Fig. \ref{fig:stat_explain}). This makes F more sensitive to patterns with empty space and aggregation, hence the name `free-space' function. F was also calculated using the border edge-correction method.\begin{br} G and F can be estimated with \citep{Gignoux1999}

\begin{equation}
\mathrm{\hat{G}}(w) = \frac{\#\{w_i\leq w, b_i > w \}}{\#\{b_i > w \}}
\end{equation}
 \begin{equation}
\mathrm{\hat{F}}(x) = \frac{\#\{x_i\leq x, b_i > x \}}{\#\{b_i > x \}},
\end{equation}
respectively,\end{br} where \#\{\dots\} is shorthand for the number of positions or events that satisfy the condition. For the case of CSR the expected value for G($w$), E[G($w$)], has the value, $\mathrm{E[G}(w)] = 1 - e^{-\pi \lambda w^2}$ \citep{Dale2014,feigelson_babu_2012}. E[F($x$)] is identical. For CSR G and F are equal, but departures from CSR cause these values to differ due to their sensitivities to clustering and empty space respectively. G and F may also be combined to form other measures, such as the Lieshout-Baddeley J function, $J(r) = 1-\mathrm{G}(w)/1-\mathrm{F}(x)$ \citep{feigelson_babu_2012} \begin{br}though they are not explored further here.\end{br}

\subsection{Second-Order Statistics}
Second-order tests look at the distributions of pairs of points, observing the change in probability compared to a random distribution of a star being within, or at, a distance $r$ from another star.

\begin{figure*}
\includegraphics[width=\textwidth]{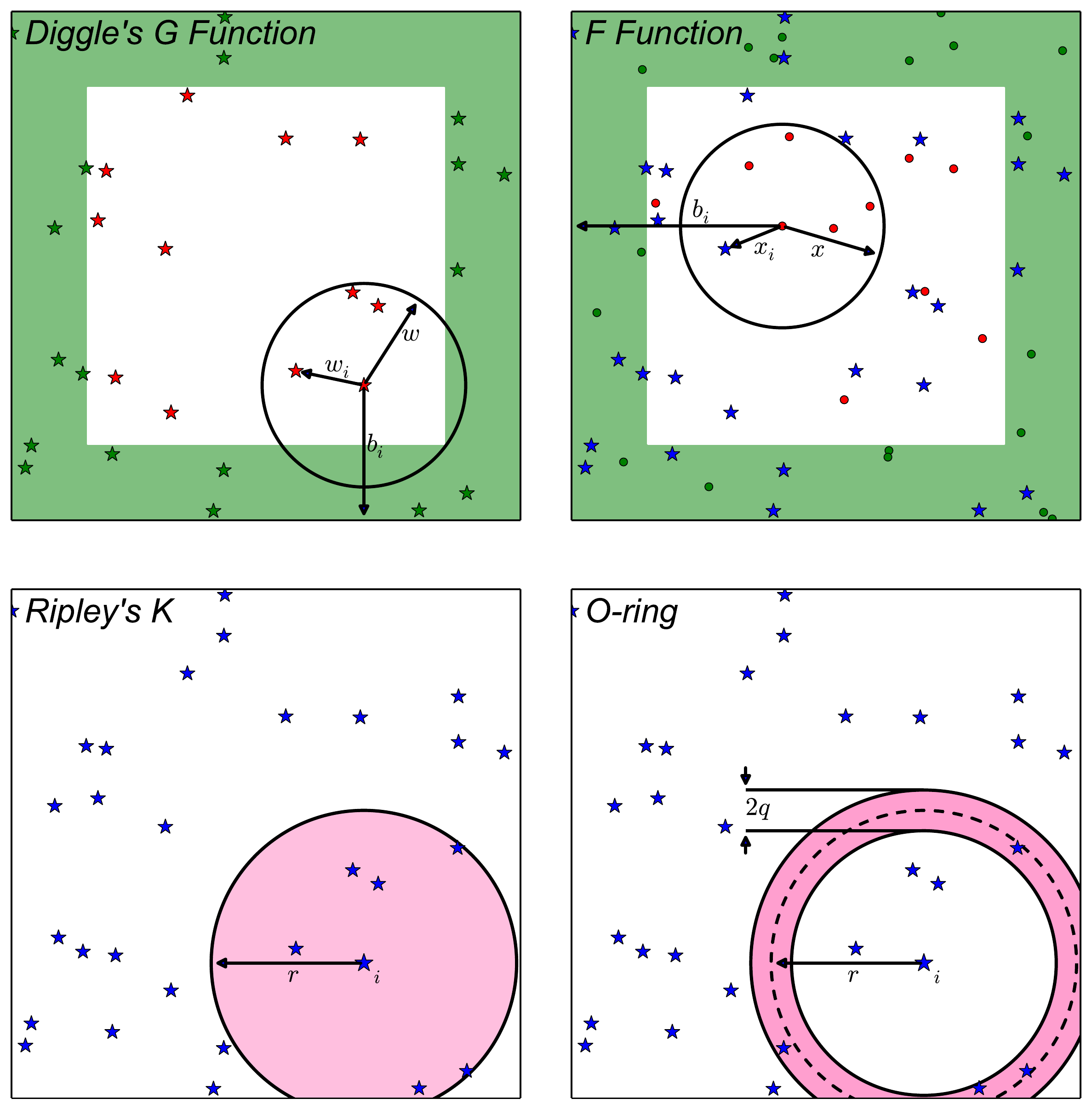}
\caption{\label{fig:stat_explain} Visual schematic of summary statistics G, F, Ripley's K and O-ring. The stars show locations of events for one realisation of CSR, the circular markers are the arbitrary positions used by F. The highlighted border in G and F shows the regions where stars/positions will be excluded by the edge-correction method. The shaded areas in Ripley's K and O-ring show the regions where $I_r(j)=1$.}
\end{figure*}

Ripley's K is one of the most commonly used spatial statistics. 
Multiplying K by the first-order intensity of the point pattern gives the expected number of events within the distance $r$ of an arbitrary event excluding the central event \citep{Wiegand2004}, $\lambda\mathrm{K}(r) = \mathrm{E}[\#\{ \text{points in area } \s\}]$, where $\s$ is a circle of radius $r$ centred on an arbitrary event. Numerically K can be estimated with \citep{Dale2014}
\begin{equation}
\mathrm{\hat{K}}(r) = \frac{|\A|}{N^2}\sum\limits_{\substack{i=1 \\ i \neq j}}^n\sum\limits_{\substack{j=1 \\ j \neq 1}}^n h_i(r)I_r(i,j),
\end{equation}
where $N$ is the number of stars, $|\A|$ is the area of the test region $\A$, $h_i(r)$ is a weighting allocated to each event for edge-correction purposes and $I_r(i,j)$ is a selection function taking the value of 1 if $d_{ij} \leq r$ and 0 otherwise. $d_{ij}$ is the distance between points $i$ and $j$. In this case $h_i = |\mathit{s}|/|\A\cap s_i|$, the inverse of the proportion of the circle $s_i$ around event \textit{i} that intersects with $\A$. This area can be calculated algebraically for rectangular windows and computationally for arbitrarily shaped regions. Fig. \ref{fig:stat_explain} presents ($s_i \cap \A$) as a highlighted area around point $i$, and points inside the highlighted area satisfy $I_r(i,j)$. $\hat{\mathrm{K}}(r)$ is not visually intuitive, however the difference between $\hat{\mathrm{K}}(r)$ and the expected value if the pattern was CSR, $\pi r^2$, gives a statistic that, while it does not contain more information, reduces the variance and is more visually intuitive \citep{Dale2014},

\begin{equation} \label{eq:L}
\mathrm{\hat{L}}(r) = \sqrt{\mathrm{\hat{K}}(r)/\pi} - r.
\end{equation}
For this convention, positive values of L$(r)$ imply clustering/aggregation and negative values imply overdispersion of the points. For the rest of the paper the L function will be used for plotting results, however the results will be discussed under the name Ripley's K. 

The O-ring statistic, sometimes referred to as the neighbourhood density function \citep{Perry2006} or the mean surface density of companions (MSDC) \citep{Larson1995}, is similar to K except it is calculated using annuli instead of circles. \begin{br}The statistic is a measure of the average density that would be observed at a distance $r$ from a star \citep{Wiegand2004}\end{br}, 
\begin{equation}
\label{eqn:oring}
\mathrm{\hat{O}}(r) = \frac{|\A|}{N^2}\sum\limits_{\substack{i=1 \\ i \neq j}}^n\sum\limits_{\substack{j=1 \\ j \neq 1}}^n h_i(r)I_r(i,j)
\end{equation}
where $I_r(i,j)$ is a selection function, taking the value of 1 if $r - q \leq d_{ij} \leq r+q$, with $q$ being the half-width of the annulus (see Fig. \ref{fig:stat_explain}), and $h_i$ is a weighting for edge correction. $h_i$ is the inverse of the proportion of the area of the annulus that lies within the boundaries. \begin{br}The use of annuli allows the values for $\mathrm{\hat{O}}(r)$ to be uncorrelated as long as the separation of values of $r$ is greater than $2\times q$, which in turn allows for an analytical approach when creating confidence envelopes (see Section \ref{sec:significance}). For CSR E$[\mathrm{\hat{O}}(r)] = \lambda$, so values larger than $\lambda$ indicate a greater-than-average density at that distance and vice-versa. The width of the O-ring annulus, $2q$, introduces the problem of binning. One rule of thumb is to begin with 
\begin{equation}
q = \rho/\sqrt{\lambda}
\label{eqn:lambda}
\end{equation}
with $\rho$ taking values between 0.1-0.2 \citep{Law2009,Yongtao_2006} and change as appropriate to maximise smoothing of the data while minimising loss of information.\end{br}

\section{SIGNIFICANCE TESTING}
\label{sec:significance}

Confidence envelopes are used to test for the significance of an individual measurement of G, F, K or O-ring. If the observed measurement exceeds the envelope the null hypothesis can be rejected with a predetermined significance $\alpha$. \begin{br}Each test in Section \ref{sec:Methods} returns a value at each considered radial distance and the spatial scale(s) on which the confidence envelope is exceeded are the scales on which the pattern is inconsistent with CSR.\end{br} In this way confidence envelopes are able to test each spatial scale against the null hypothesis while maintaining the spatial information that is present.

\subsection{Confidence Envelopes}
The expected distributions of these test statistics are typically not known, and so to test the significance of an observed statistic, $T_1(r)$, they are compared to a global confidence envelope. \begin{br}A pointwise envelope tests if $T_1(r)$ is among the $k^{th}$ most extreme values among the set of $T_i(r)$ values for i = 1,\ldots , \textit{n}+1, where \textit{n} is the number of simulated patterns for a null hypothesis $H_0$. A pointwise envelope can then reject a null hypothesis with probability $2k/(n+1)$ at a single distance scale $r$ if the envelope is exceeded. This probability is valid when testing a single distance scale; if this is used for multiple scales the probability of rejection is increased. Visually inspecting where the observed statistic exceeds the pointwise envelopes is a simultaneous test across all values of $r$ and therefore does not have the expected significance level.
 
A global confidence envelope is one which allows simultaneous testing of all probed radial distances at a predetermined significance level, meaning that the graphical plot of $T_1(r)$ among the envelopes is a valid statistical tool for testing hypotheses,\end{br} as well as looking for scales at which $H_0$ is rejected. The global envelopes utilised in this paper are \textit{directional quantile maximum absolute difference} (MAD) envelope tests, following the description from \citet{myllmaki2017}, where the upper and lower bounds are of the form

\begin{align}
T^u_{\mathrm{low}}(r) &= T_0(r) - u|\underline{T}(r) - T_0(r)|,\\
T^u_{\mathrm{upp}}(r) &= T_0(r) - u|\overline{T}(r) - T_0(r)|
\end{align}

where $T_0$ is the expected value under the null hypothesis, $\overline{T}(r)$ and $\underline{T}(r)$ are the 2.5 per cent upper and lower quantiles of the distribution of $T(r)$ under $H_0$, and $u$ is a parameter which determines the confidence level of the envelopes. The $u$ values are the maximum deviation of each $T_i(r)$ from $T_0(r)$ and scaled by $|\overline{T}(r) - T_0(r)|$ and $|\underline{T}(r) - T_0(r)|$ for values above and below $T_0$ respectively, and for a significance level of $\alpha$ the $\alpha (n+1)$th largest value among the set of $u$ values is used.

\begin{figure}
\includegraphics[width=\columnwidth]{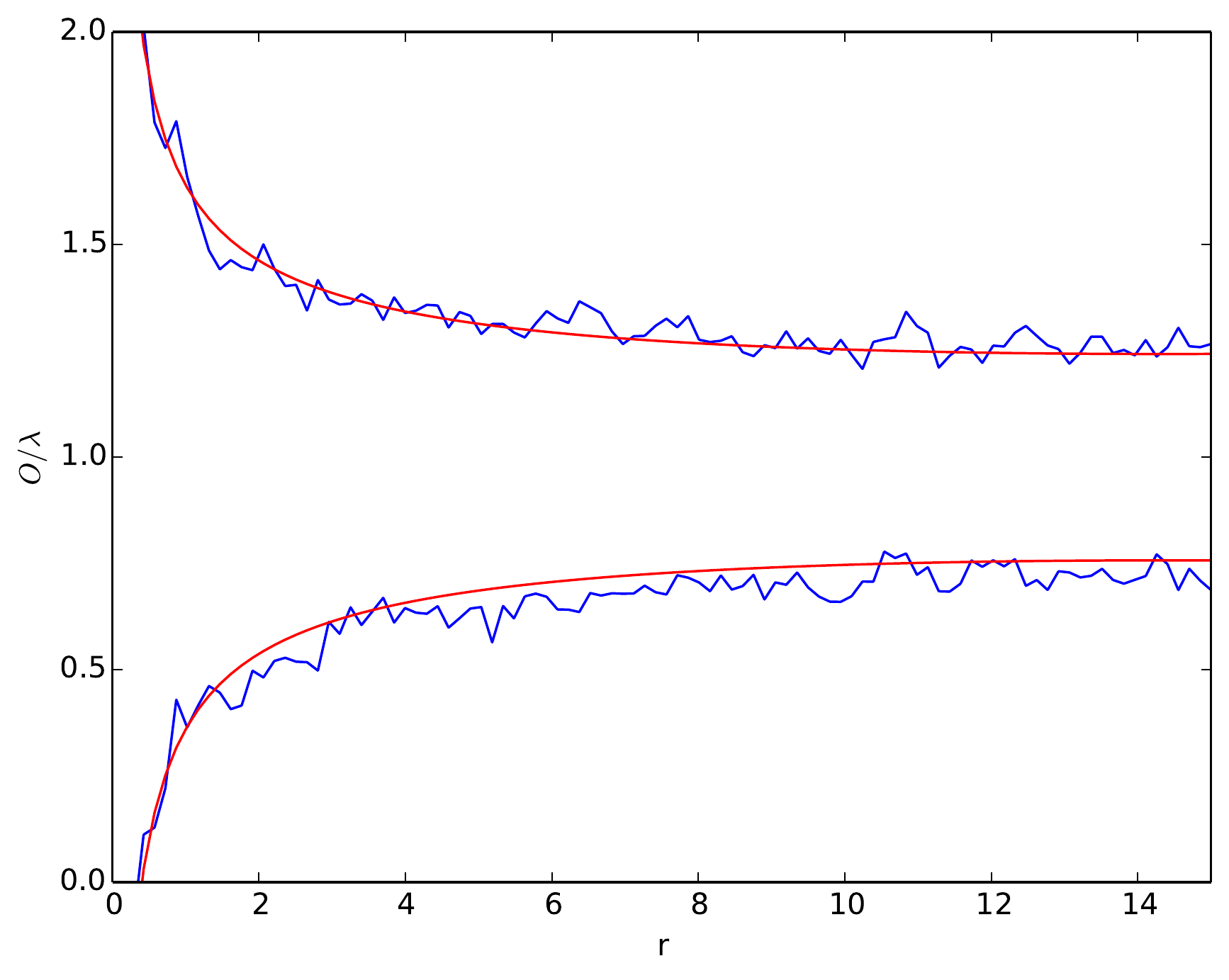}
\caption{\label{fig:analytical envelopes} Comparison of analytical (red) vs empirical Monte Carlo (blue) confidence envelopes for O-ring for CSR with $\lambda = 0.133$ in a 30x30 arb. units area.}
\end{figure}

\begin{br2}It is possible to produce global confidence envelopes analytically without requiring simulation when the distribution of a statistic is known. From the analytical perspective the width of a confidence envelope is a representation of the variance of the statistic under a given null hypothesis. This variance is a function of factors including the number of stars and the area and geometry of the study window \citep{wiegand2016}. With increasing numbers of stars the width of the envelopes is decreased and the statistics gain an increase in power while low numbers can make rejection of a null hypothesis unlikely \citep{Gignoux1999}. Analytical envelopes function best when the tests for each $r$ are independent, which for O-ring requires that each $r$ position be offset from the previous by at least $q$ and is not possible for Ripley's K due to the nature of the test. While these analytical envelopes were not used, they validate the empirical envelopes, as shown in Fig. \ref{fig:analytical envelopes}. \end{br2}

\subsection{Envelope examples}
\label{subsec:envelope}
\begin{figure*}
\subfloat{\includegraphics[width=0.5\linewidth]{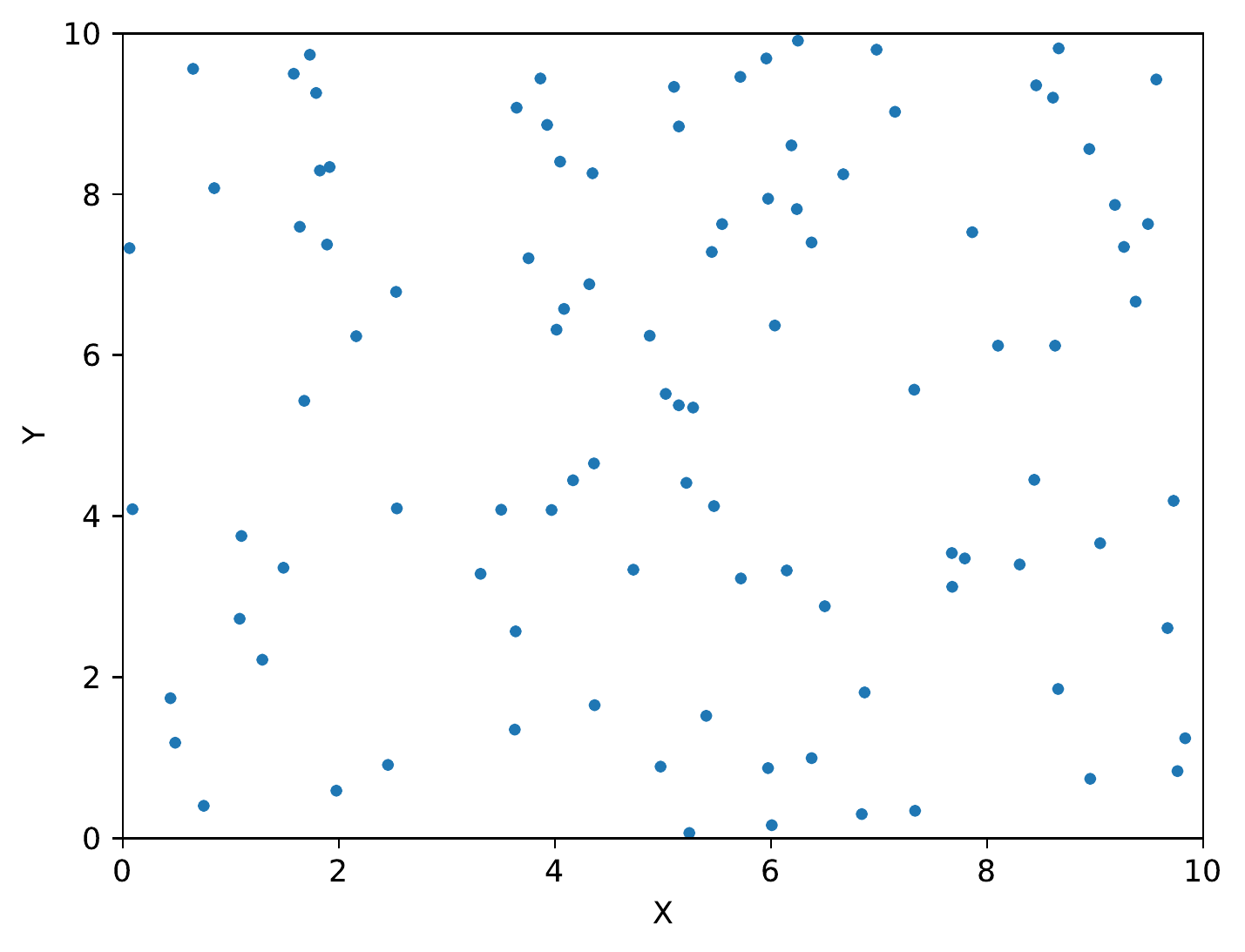}}
\subfloat{\includegraphics[width=0.5\linewidth]{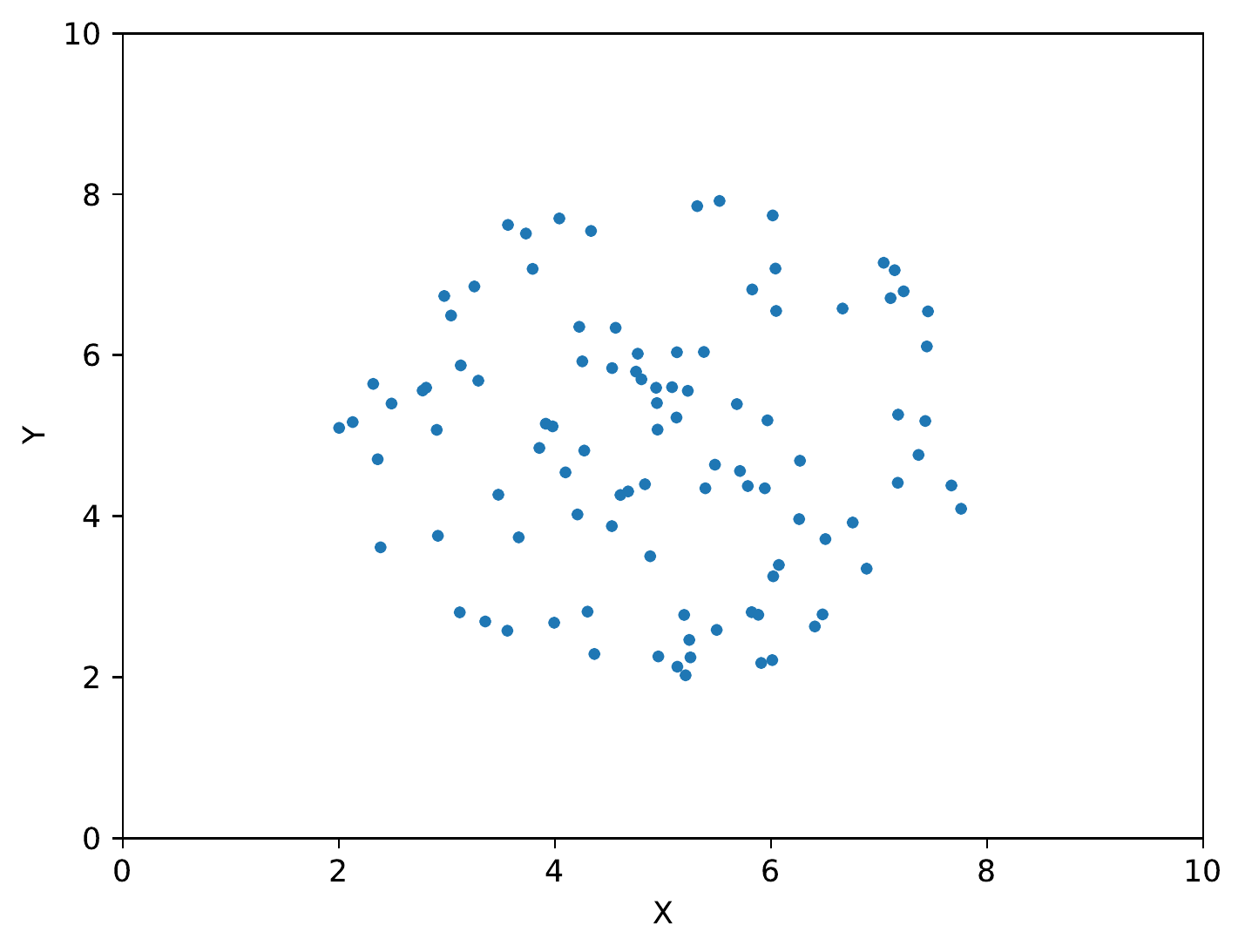}}\\
\subfloat{\includegraphics[width=0.5\linewidth]{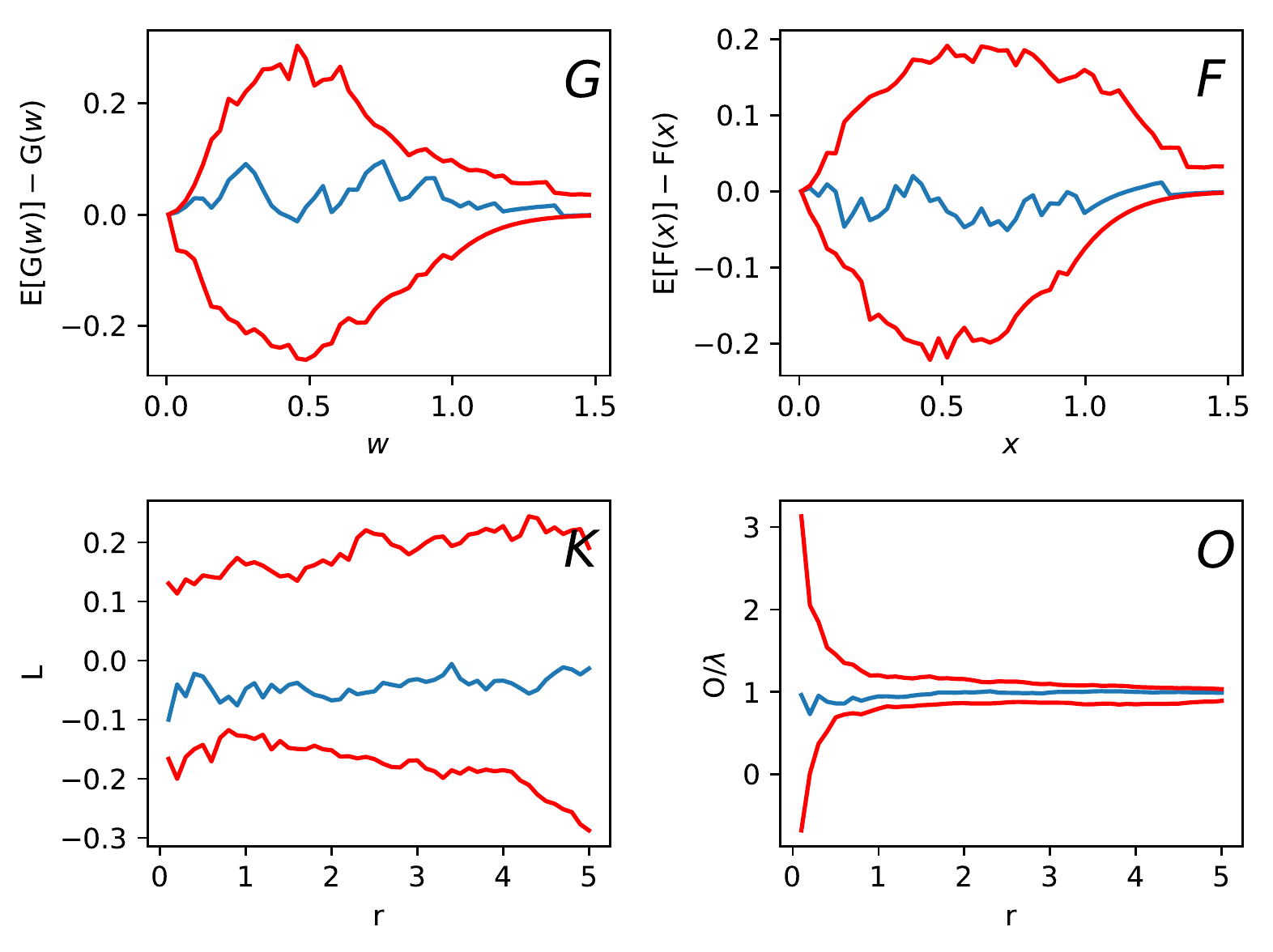}}
\subfloat{\includegraphics[width=0.5\linewidth]{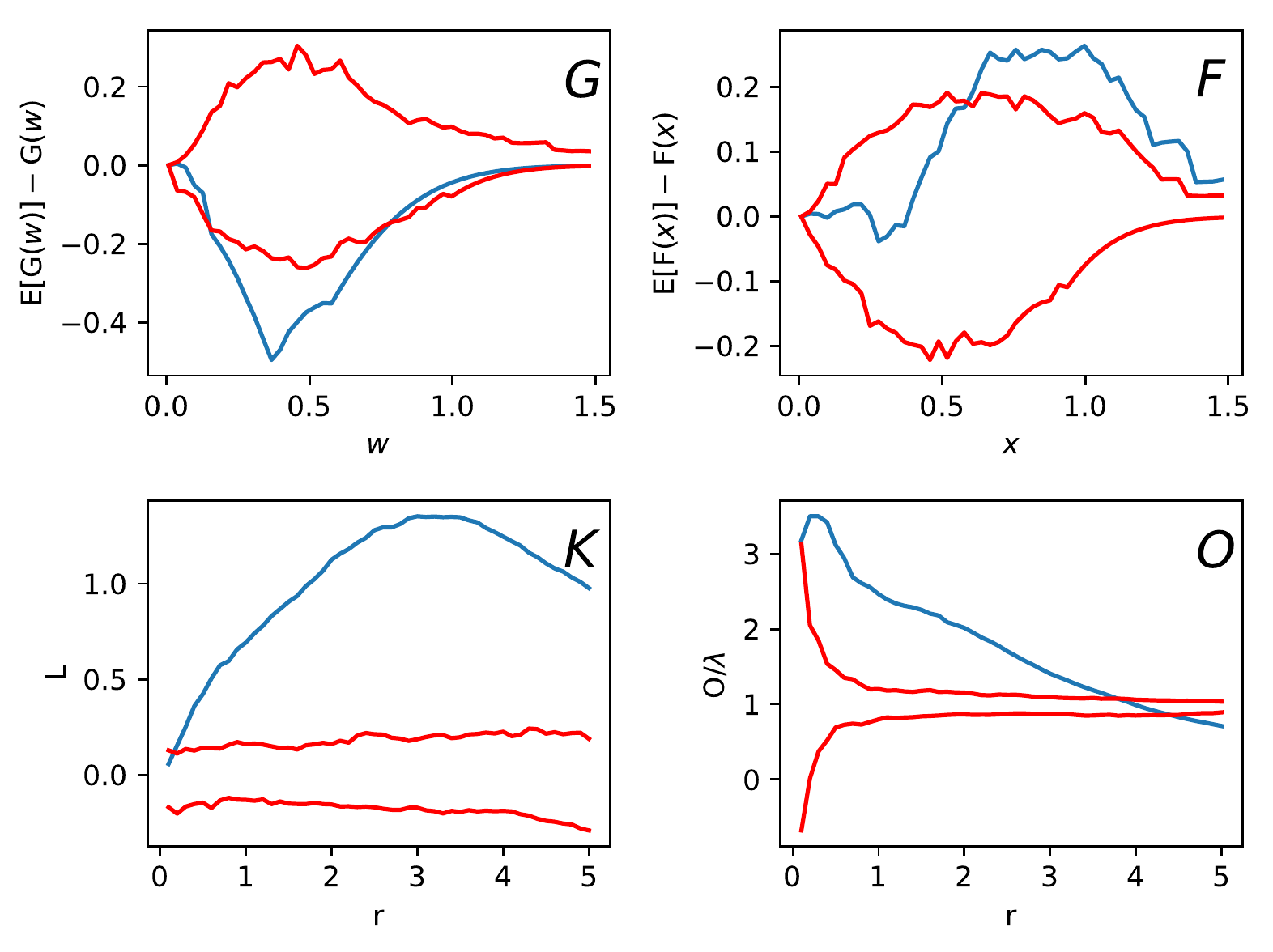}}\\
\caption{Left: (above) Realisation of CSR with $\lambda = 1.0$, (below) results of G, F, K and O-ring in blue with 95 per cent global confidence envelopes for CSR in red. Right: (above) Realisation of centralised cluster with $\lambda = 1.0$ and $R=3$, (below) G, F, K and O-ring results in blue with 95 per cent global confidence envelope for CSR in red.}
\label{fig:csr_and_cluster_examples}
\end{figure*}

\begin{figure}
\subfloat{\includegraphics[width=\columnwidth]{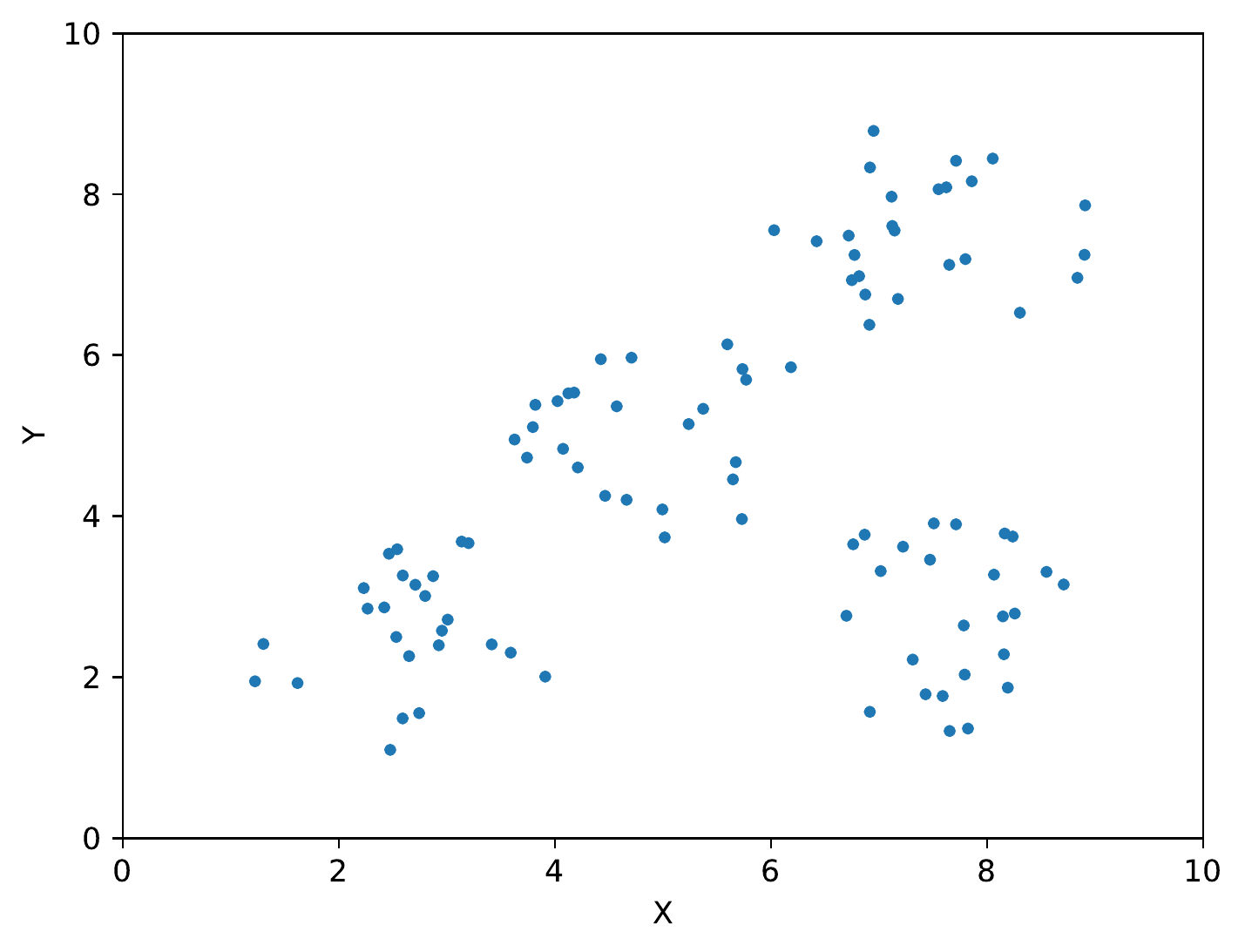}}\\
\subfloat{\includegraphics[width=\columnwidth]{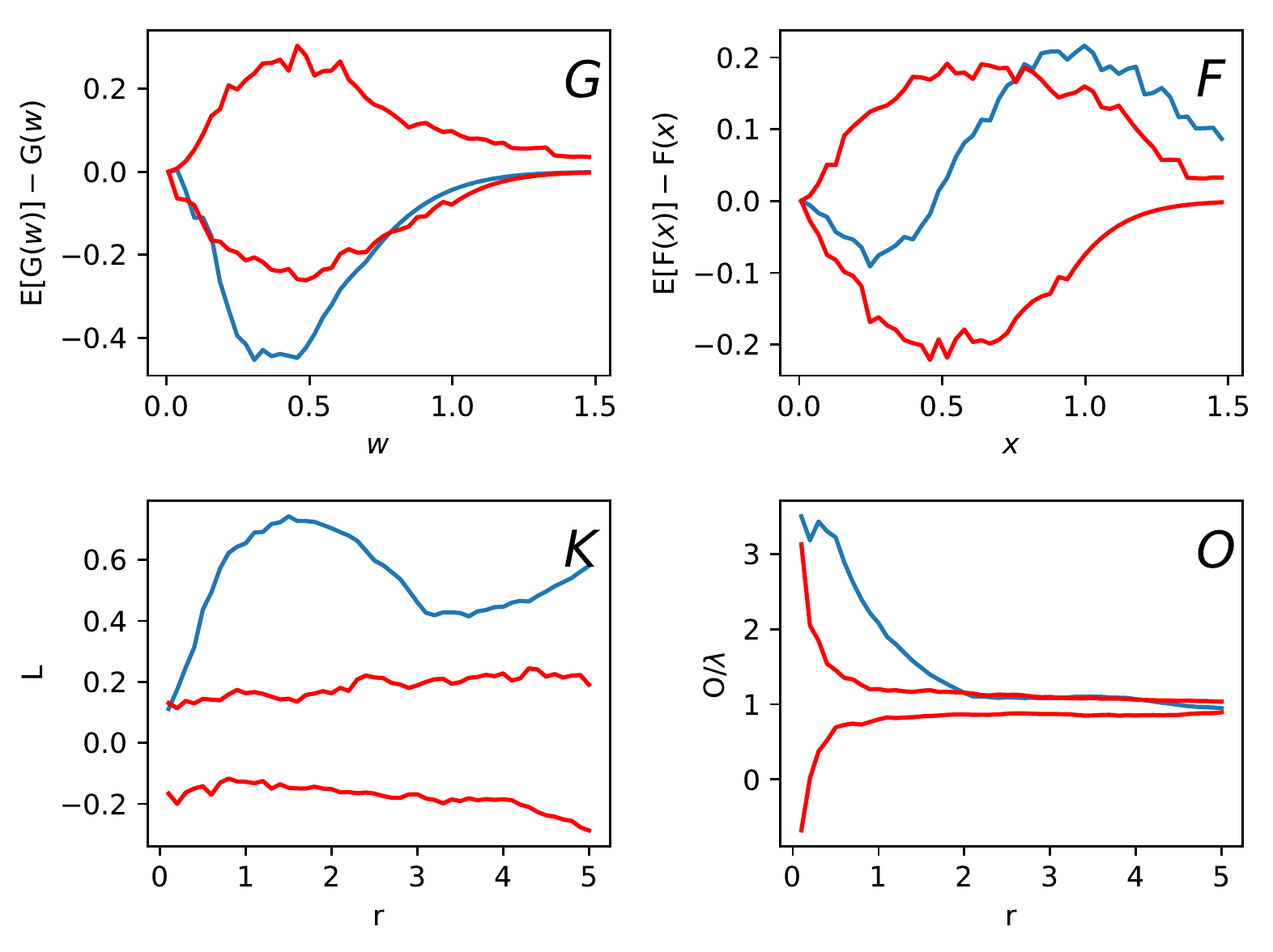}}\\
\caption{(above) Realisation of 4 clusters with $\lambda = 1.0$ and $R=1.5$, (below) G, F, K and O-ring results in blue with 95 per cent global confidence envelope for CSR in red.}
\label{fig:multicluster_example}
\end{figure}

\begin{br}Figs. \ref{fig:csr_and_cluster_examples} and \ref{fig:multicluster_example} show three examples of spatial point processes, along with the results from the methods described above. The left-hand panel of Fig. \ref{fig:csr_and_cluster_examples} shows a realisation of CSR with $\lambda = 1$ keeping the number of points constant. In this example, all of the statistics agree that CSR cannot be rejected as a null hypothesis for the spatial point process as they remain within the confidence envelope. The right-hand panel shows a single centralised cluster produced from a uniform circular probability density function with radius $R=3$. All four statistics reject CSR for this pattern. Fig. \ref{fig:multicluster_example} contains a realisation of four clusters each produced with a circular probability density function of radius $R=1.5$. For the four clusters every statistic rejects CSR with O-ring and K demonstrating behaviours due to the presence of multiple clusters.\end{br}

K and O-ring are used to study the second-order effects of the spatial point process; however, the cluster examples in Figs. \ref{fig:csr_and_cluster_examples} and \ref{fig:multicluster_example} do not contain any second-order effects as the star positions are drawn from a non-homogeneous uniform probability distribution. The deviations are therefore due to large first-order effects which give the impression of clustering due to second-order effects, a phenomenon known as virtual aggregation \citep{Wiegand2004}. As such, when the rejection of CSR using O-ring and K are discussed the nature of the effect (either first or second-order) will not be commented on, only that the measured pattern is consistent or inconsistent with CSR.

Excursions from the envelope represent spatial scales at which the null hypothesis used to produce the envelope can be rejected. It is important, however, to understand what the tests are measuring to be able to interpret these regions. O-ring is a measure of the average density we would observe at a distance $r$ from a star, therefore an excursion at a given scale is indicating a significantly over- or under-populated region. K (or L), being a cumulative statistic, is measured up to a given radial distance and describes up to which scales a pattern rejects the null hypothesis of the envelope. For this reason O-ring is typically easier to interpret though both statistics contain the same information, as O-ring is related to the differential of K. \begin{br2} The differences between O-ring and K can be clearly seen in the right-hand panel of Fig. \ref{fig:csr_and_cluster_examples} at $r = 4$ where O-ring appears to be consistent with CSR while K rejects the null hypothesis. Therefore, any scales indicated by excursions are both a function of the statistic and the null hypothesis represented by the envelope. \end{br2}

\section{TRIALS}
\label{sec:trials}
We tested the four statistics for their ability to reject CSR for a single cluster in the presence of background noise. To set up this experiment, a cluster with population $\Nc$ was projected on to a field of background stars of number $\Nbg$. The positions of the cluster population were generated by sampling a two-dimensional Gaussian probability density function with Gaussian width $\sigma =R/2$ centred on the midpoint of the study window, where $R$ is a characteristic radius chosen for the cluster. The background population was $\Nbg$ sets of (x,y) coordinates randomly distributed across the study window. 

\begin{br}Multiple cluster pattern realisations of each combination of $\Nc$ and $\Nbg$ were tested and their summary statistics compared to 95 per cent confidence envelopes for a null hypothesis of CSR. The fraction of realisations that reject CSR are then a measure of the likelihood that a given pattern with $\Nc$ cluster members and $\Nbg$ background points will be correctly identified as non-random. The envelopes were generated using the total number of points, \br{$\Ntot = \Nc + \Nbg$}, randomly distributed across the study window for each CSR realisation.\end{br}

For an order of magnitude indication of whether the tests were functioning correctly a signal-to-noise calculation was performed for the study region as whole. By assuming the number of clustered stars can be estimated by subtracting an estimation of the background population, a possible measure of the SNR for the study window is
\begin{equation}
\label{eqn:SNR}
\mathrm{SNR} = \frac{N_c}{\sqrt{N_c+N_{bg}}}.
\end{equation}
The expected behaviour for a given $\Nc$ should then be a decrease in cluster realisations that reject CSR with decreasing SNR.

\section{RESULTS}
\label{sec:Results}
The following results show the effectiveness of these statistical tests in rejecting CSR when the number of cluster points, $\Nc$, and the total number of points in the window, $\Ntot$ are controlled separately. 
To produce the confidence envelopes 199 realisations of CSR were generated for each value of $\Ntot$. To test the different numbers of cluster points and background points 30 simulations of each ($\Nc$, $\Ntot$) combination were generated.
The fraction of simulated observations which reject CSR, from here on referred to as the rejection fraction, is an estimate of the empirical probability that a given cluster of $\Nc$ members projected in-situ with $\Nbg$ background members would be identified as non-random. As a metric to compare the different statistics, the total number of detections was divided by the total number of trials run within a range of $\Nc$ and $\Ntot$ values and presented as the detectability score. The detectability score can take values from 0 to 1, representing the extremes of no trials rejected and all trials rejected respectively.
For G, F and Ripley's K, the detectability scores were calculated from 10 unique computations of the entire $\Nc$ and $\Ntot$ parameter space; for each variation in the annulus width for O-ring only one computation was produced and so uncertainties are not available for those values.

\subsection{Diggle's G and the `empty space' function}
\begin{figure}
\centering
\includegraphics[width=\columnwidth]{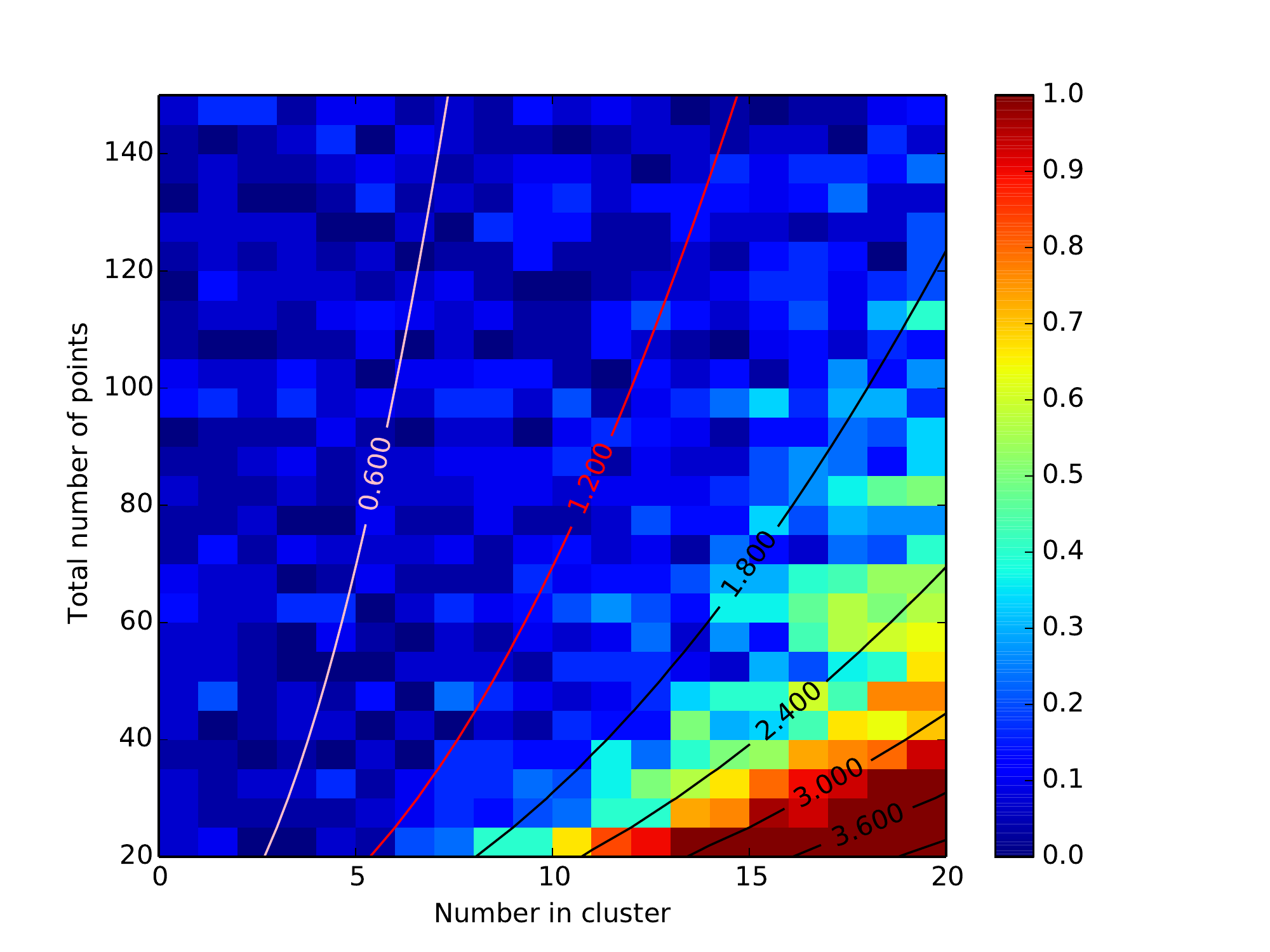}
\caption{The rejection fraction with $P(H_0)<5\%$ for Diggle’s G function. The contours show the theoretical SNR (Eqn. \ref{eqn:SNR})}
\label{fig:resultsG}
\end{figure}

\begin{figure}
\centering
\includegraphics[width=\columnwidth]{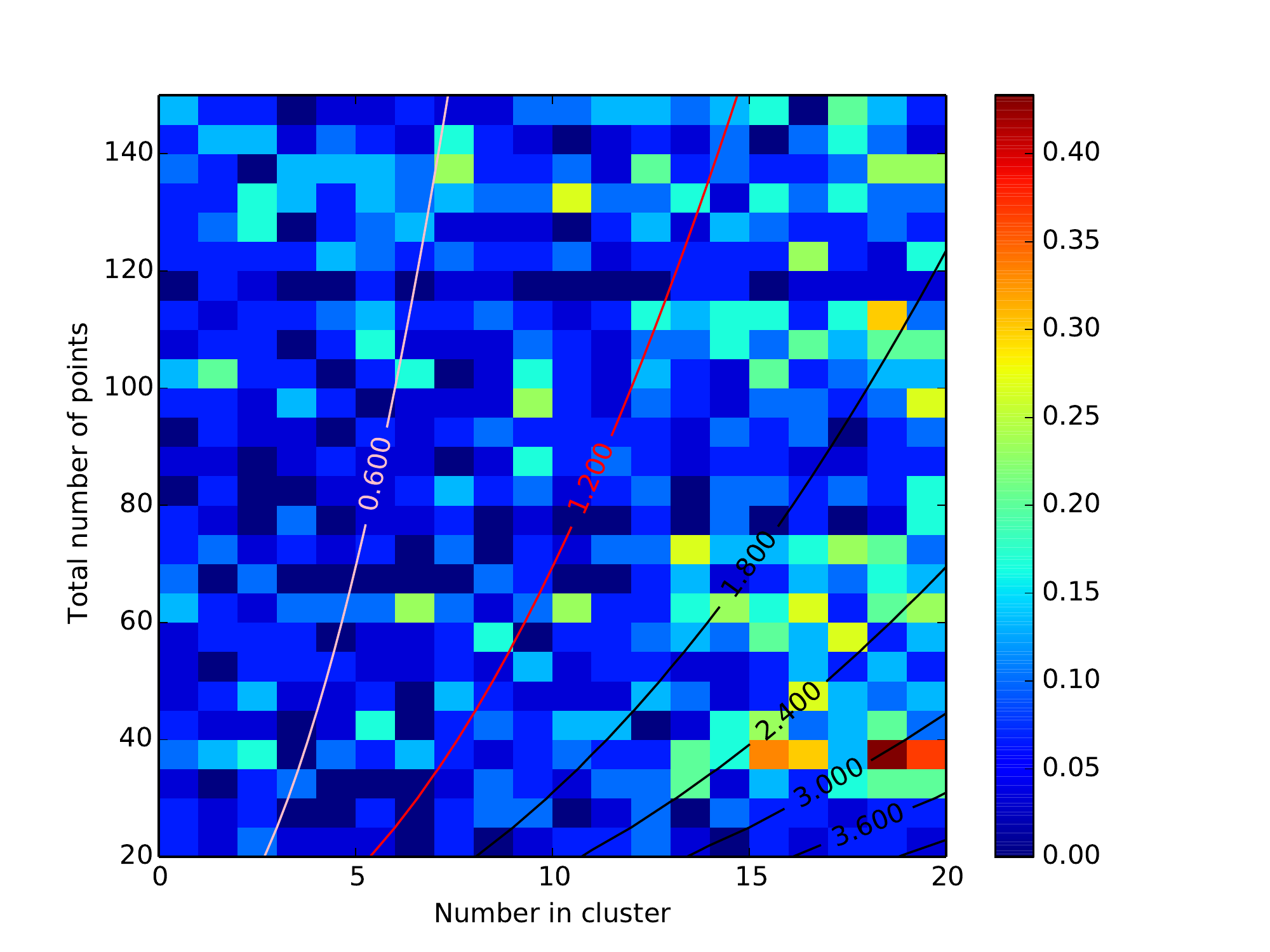}
\caption{The rejection fraction with $P(H_0)<5\%$) for F. The contours show the theoretical SNR (Eqn. \ref{eqn:SNR}). Note that the colour scale is over a much reduced range.}
\label{fig:resultsF}
\end{figure}

The results using E[G($w$)]-G($w$) and E[F($x$)]-F($x$) are shown in Figures \ref{fig:resultsG} and \ref{fig:resultsF} respectively. For G only the largest clusters with the least contamination consistently rejected CSR, while the results for F appear only weakly correlated with position in the parameter space. This is reflected in their detectability scores of $0.174 \pm 0.004$ and $0.089 \pm 0.001$ respectively.

\subsection{Ripley's K}
\begin{figure}
\centering
\includegraphics[width = \columnwidth]{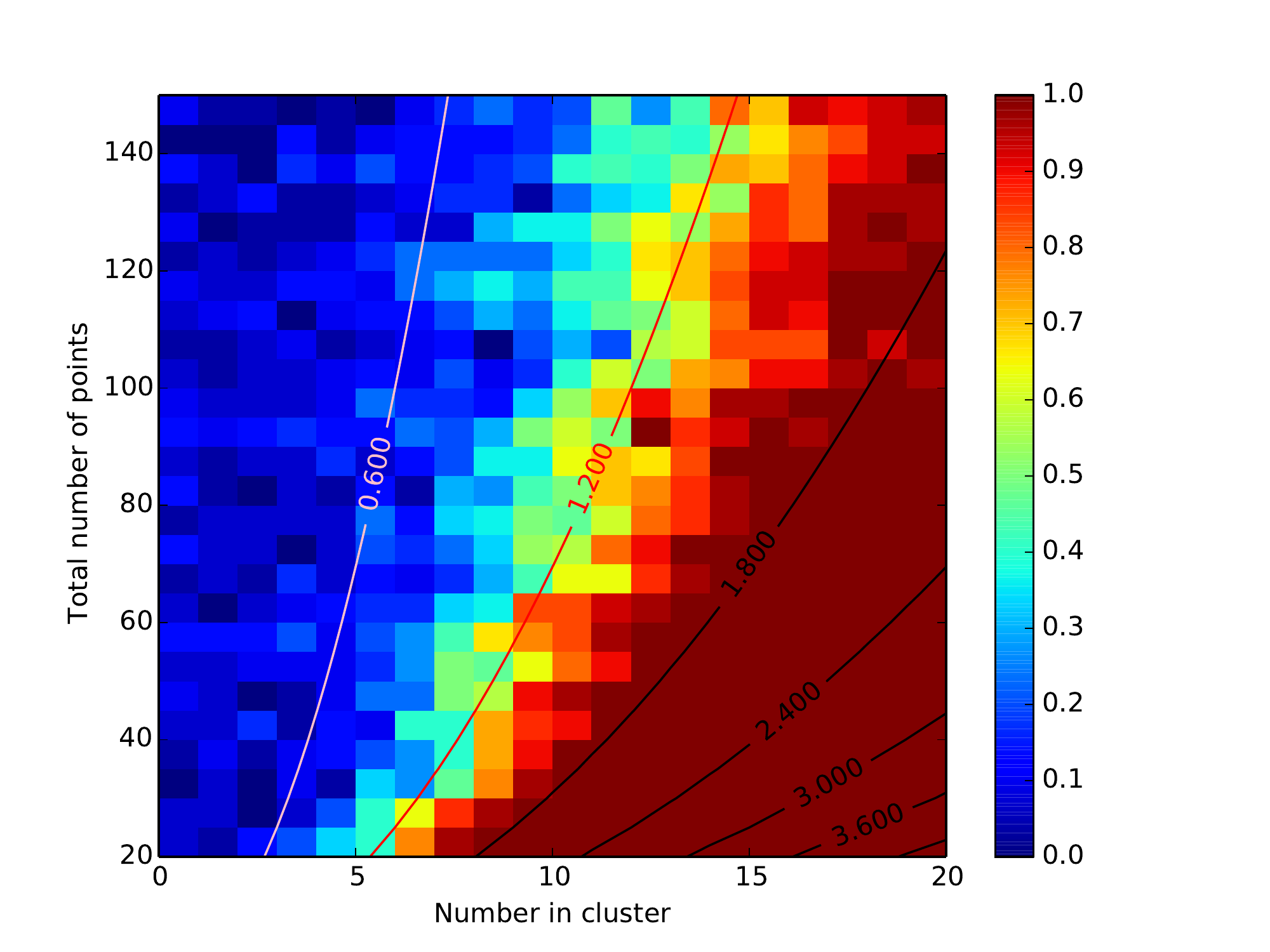}
\caption{The rejection fraction with $P(H_0)<5\%$ for Ripley's K. The contours show the theoretical SNR (Eqn. \ref{eqn:SNR})}
\label{fig:resultsK}
\end{figure}

Ripley's K rejects far more clusters than F and G as shown in Fig. \ref{fig:resultsK}. The improved rejection fraction \begin{br}demonstrates a signal-to-noise effect whereby a given cluster can be masked by an increase in background population.\end{br} The row with zero cluster members are simply runs of CSR; with a confidence envelope of 95 per cent the chance of at least one false positive is approximately 80 per cent for each value of $\Ntot$. The detectability score for Ripley's K was $0.550 \pm 0.003$.

\subsection{O-ring}
\label{sub:Oring}

\begin{figure}
\centering
\subfloat{\includegraphics[width=\columnwidth]{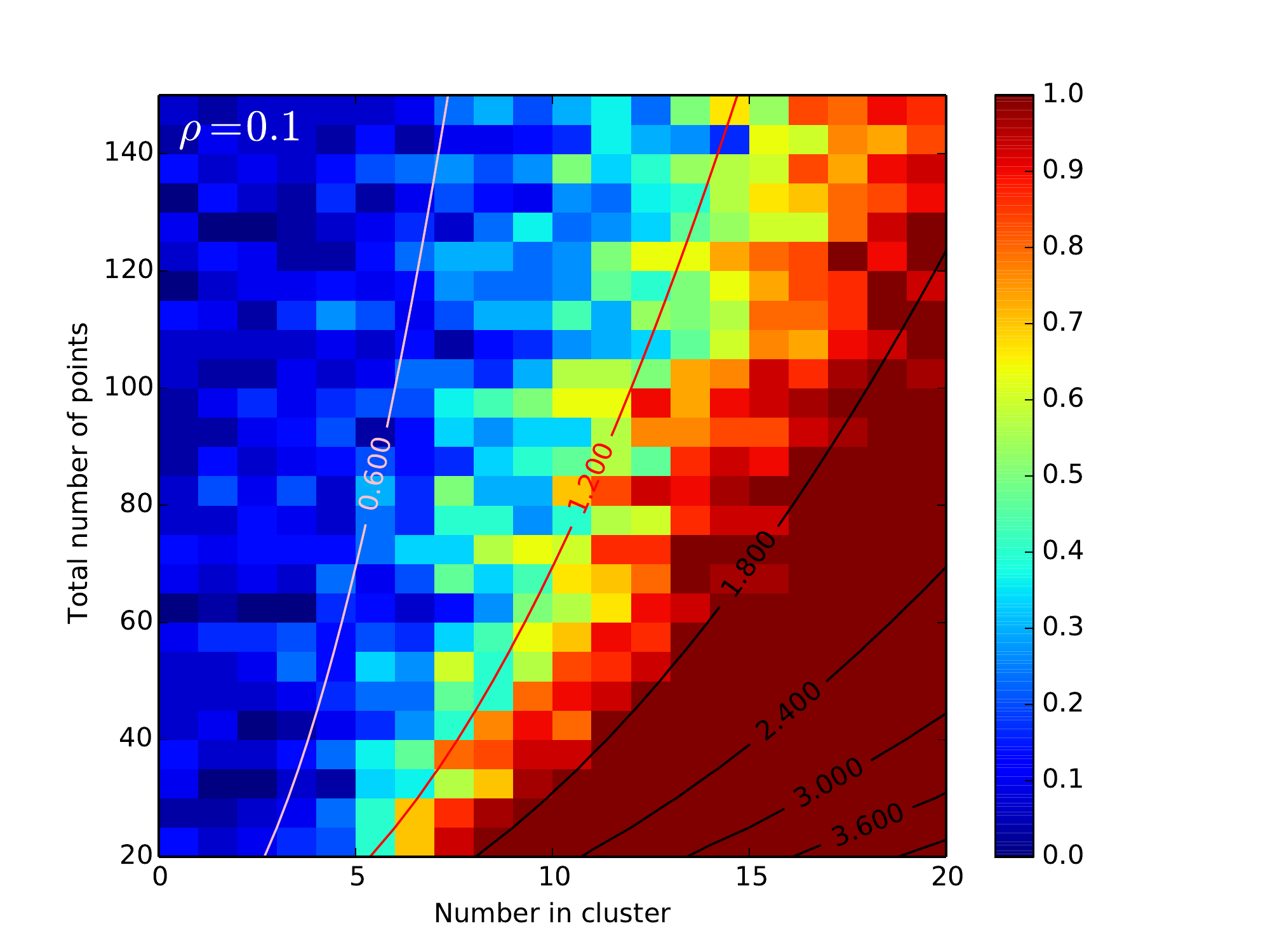}}\\
\subfloat{\includegraphics[width=\columnwidth]{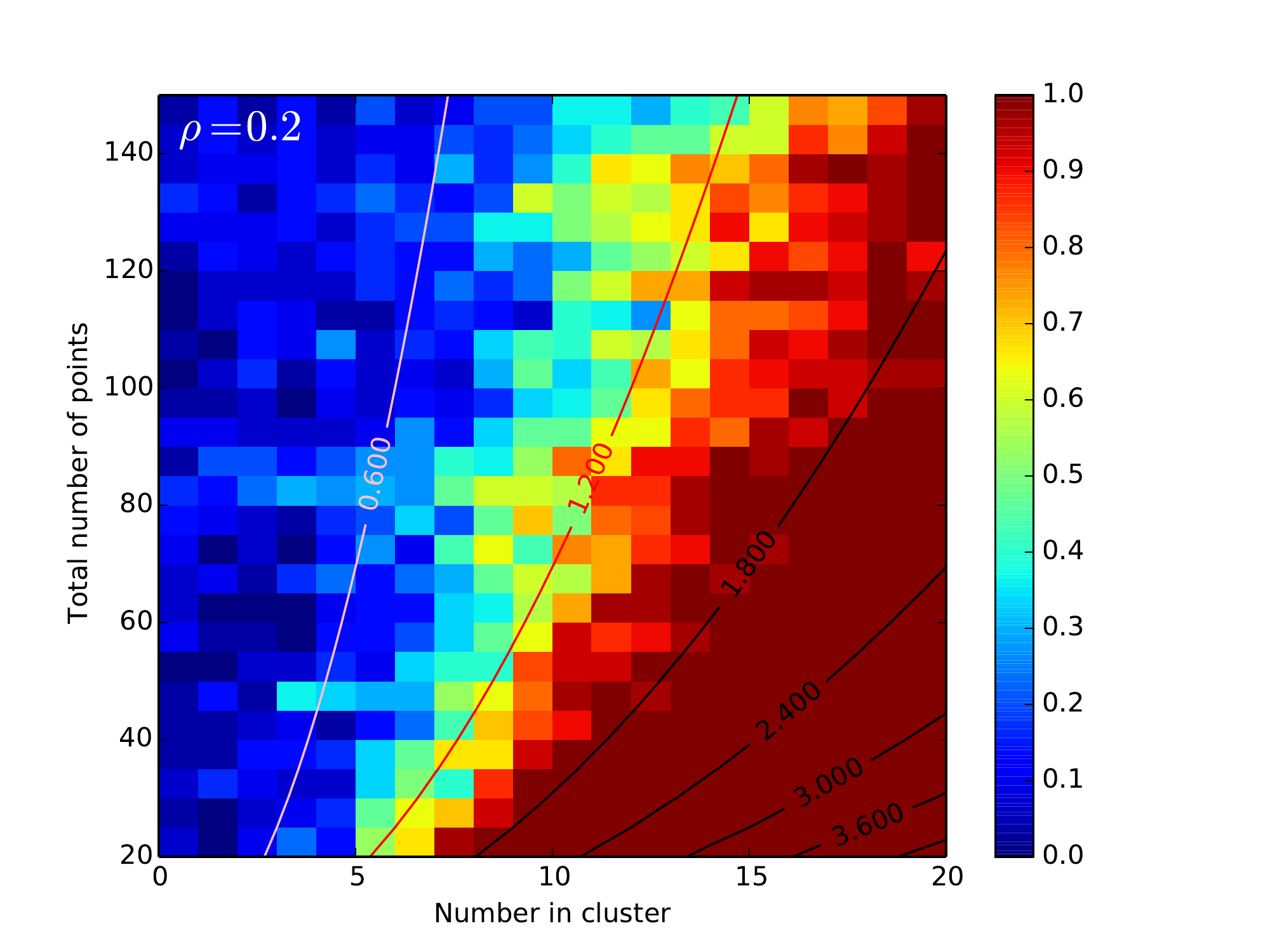}}
\caption{Rejection fraction as a function of annulus width for the O-ring statistic. Annulus widths decrease with the total number of points according to Equation \ref{eqn:lambda}. The contours show the theoretical SNR (Eqn. \ref{eqn:SNR}).}
\label{fig:varWidth}
\end{figure}

O-ring has an additional parameter compared to the other tests --- the width of the annuli. This width determines the amount of area contained within each annulus and therefore the number of cluster and non-cluster points contained within. The benefit of Equation \ref{eqn:lambda} is that the bin widths are decided without prior knowledge of the existence or scale(s) of the cluster(s) in the study window. The results of using $\rho = 0.1$ and $0.2$ in Equation \ref{eqn:lambda} are shown in Fig. \ref{fig:varWidth}. The change in detectability scores between the two values can be seen in Table \ref{table:decto_scores}.

\begin{figure*}
\subfloat{\includegraphics[width=0.5\linewidth]{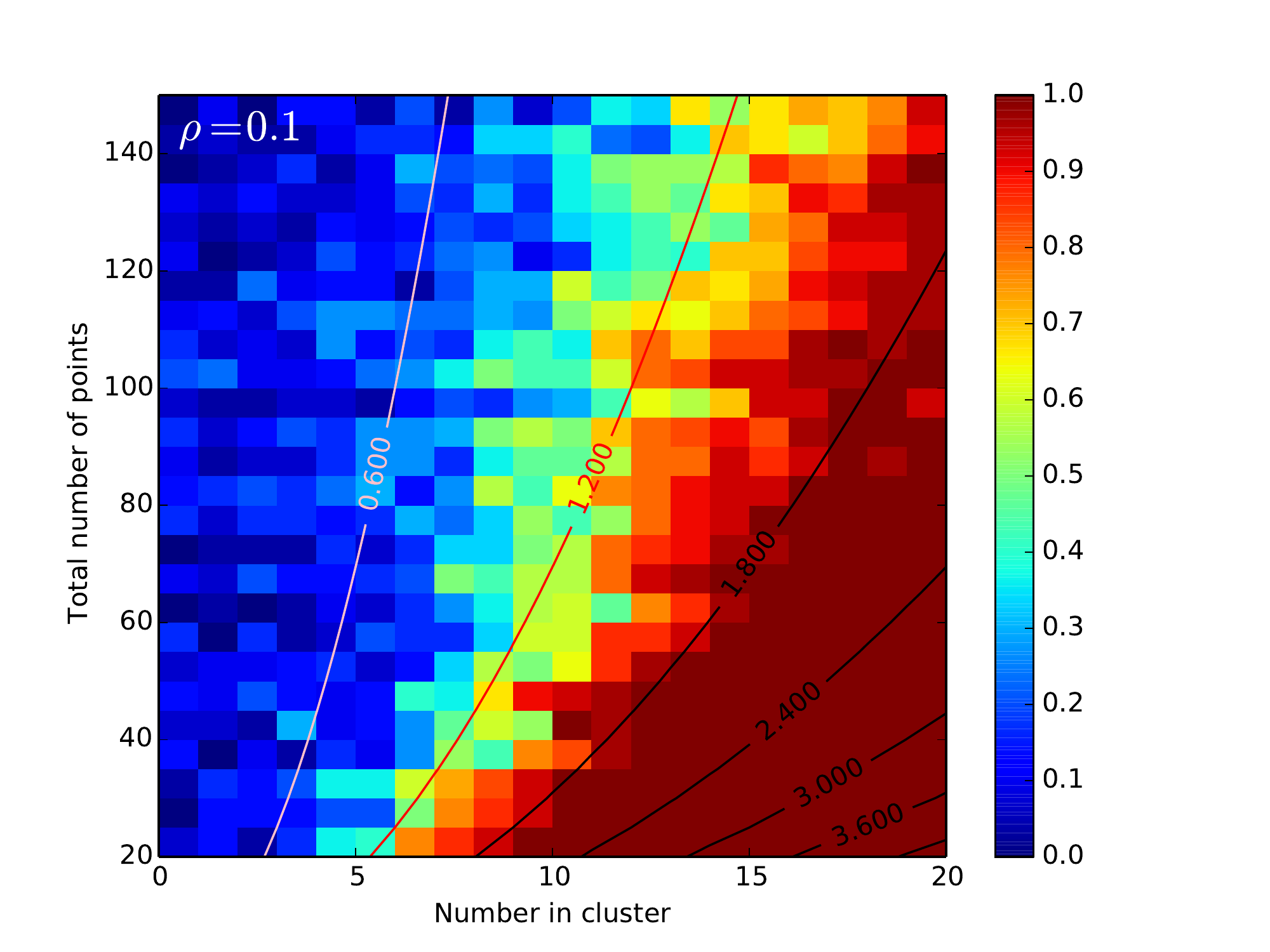}}
\subfloat{\includegraphics[width=0.5\linewidth]{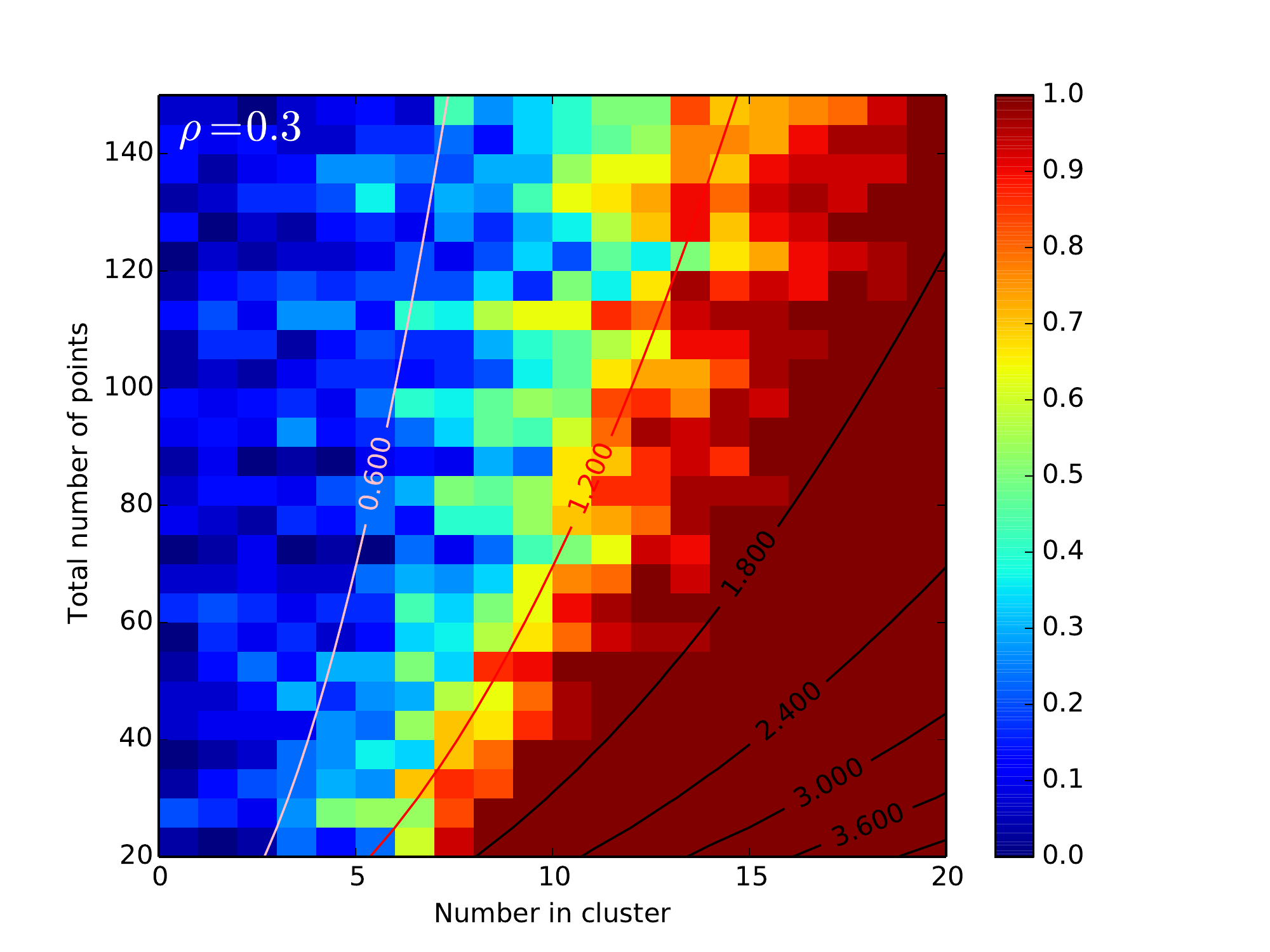}}\\
\subfloat{\includegraphics[width=0.5\linewidth]{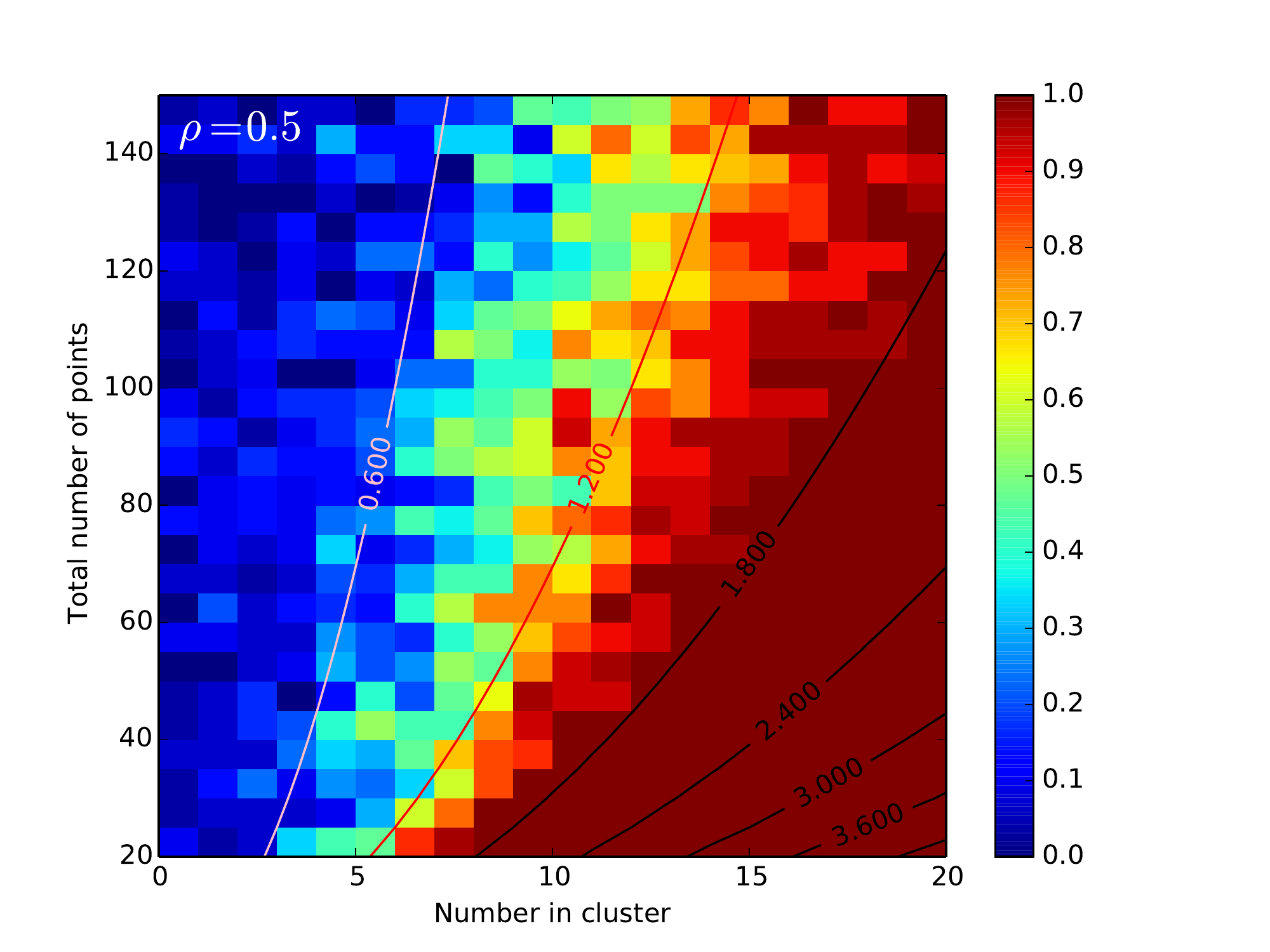}}
\subfloat{\includegraphics[width=0.5\linewidth]{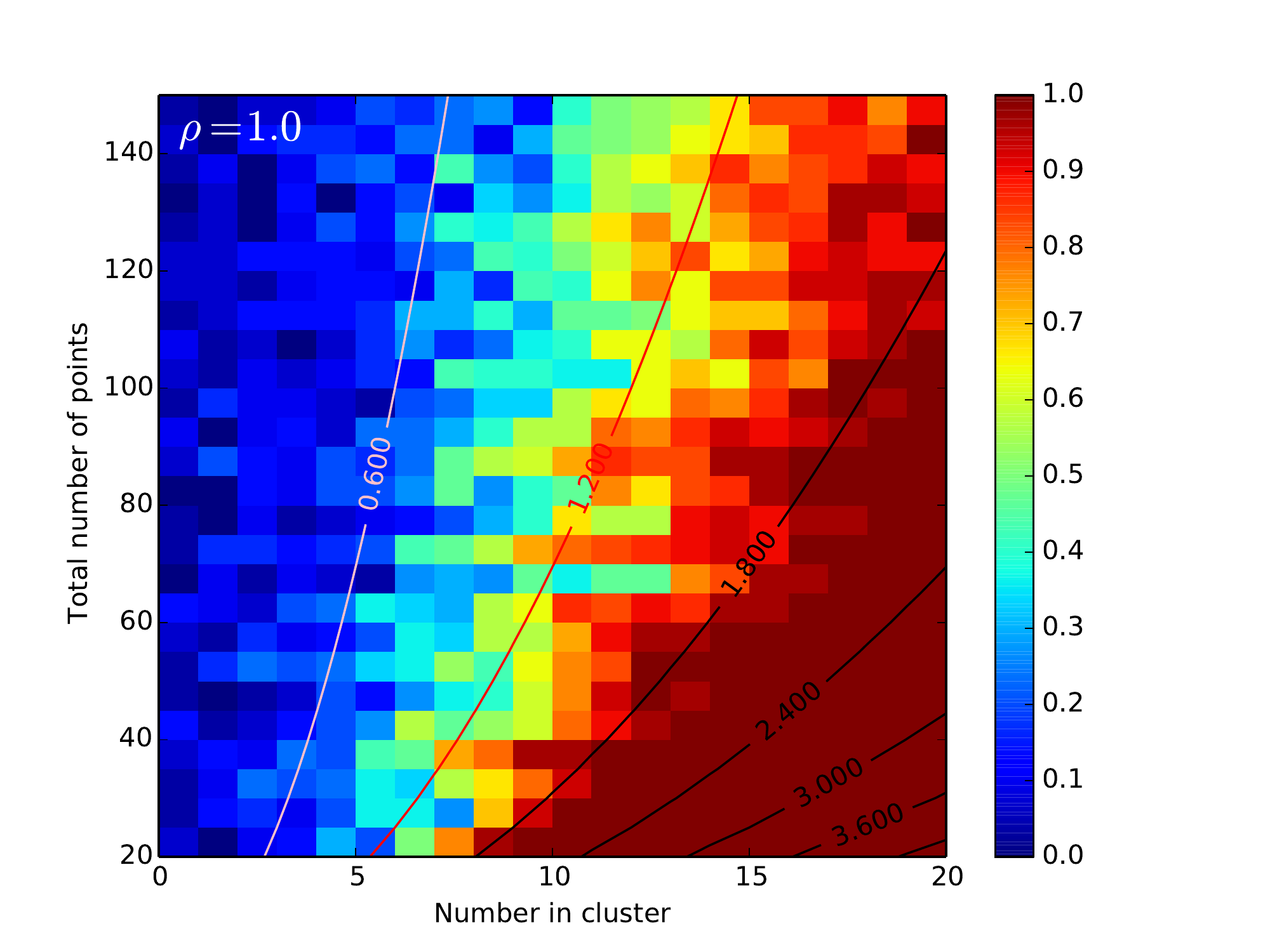}}
\caption{The rejection fraction with $P(H_0)<5\%$ for O-ring statistic with fixed annuli widths. The widths used in each parameter space were given by $\rho\times R$ with $\rho = 0.1, 0.3, 0.5 \text{ and } 1.0$ respectively and $R$ is the radius of the cluster.  The contours show the theoretical SNR (Eqn. \ref{eqn:SNR}).}
\label{fig:fixedWidth}
\end{figure*}

\begin{figure}
\includegraphics[width=\columnwidth]{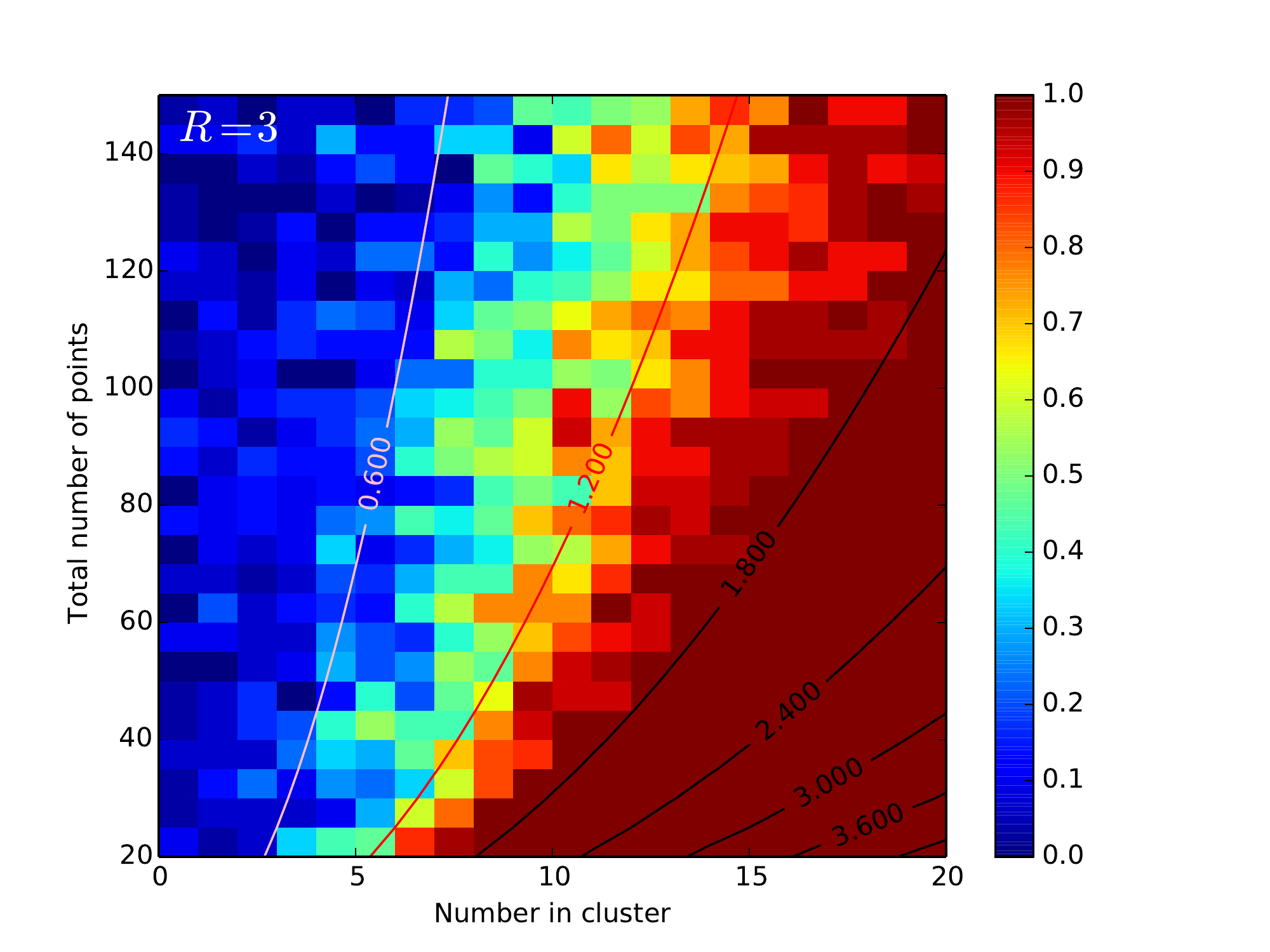}
\includegraphics[width=\columnwidth]{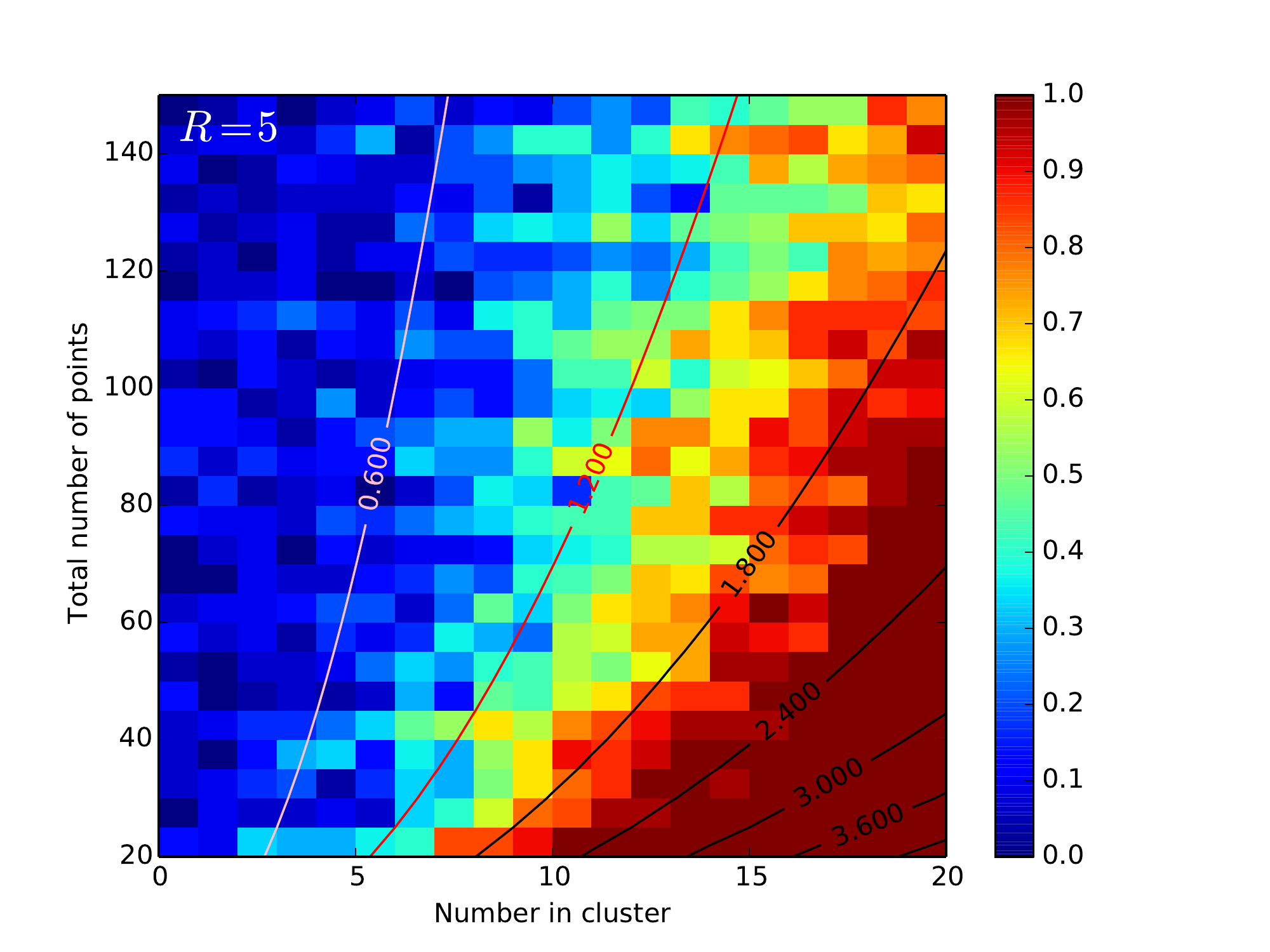}
\caption{The rejection fraction with $P(H_0)<5\%$ for O-ring statistic with fixed annuli widths for cluster radii 3 and 5.  The annulus width has been kept constant for both sets of trials, $q = 1.5$ arb. units. The contours show the theoretical SNR (Eqn. \ref{eqn:SNR}).}
\label{fig:compare3n5}
\end{figure}

Having a different annulus width for each total number of points obfuscates the direct effect of the width on the detectability fraction and so Fig. \ref{fig:fixedWidth} shows the same region of parameter space except the annulus width has been kept constant across all positions in the parameter space. Here the annulus width is some multiple of the radius of the cluster. The radius of the cluster is not a value often known \textit{a priori}, however this demonstrates that an annulus with a width larger than the radius of the cluster begins to degrade the ability of the statistic to reject CSR. As is to be expected, the rejection fraction is both a function of the degree of clustering and the annulus width. Fig. \ref{fig:compare3n5} shows the effect of keeping the annulus width constant and adjusting the characteristic radius of the cluster. \begin{br}Overall the likelihood of rejection is decreased for a more dispersed cluster, as\end{br} shown in Table \ref{table:decto_scores}.

\begin{figure*}
\subfloat{\includegraphics[width=0.5\linewidth]{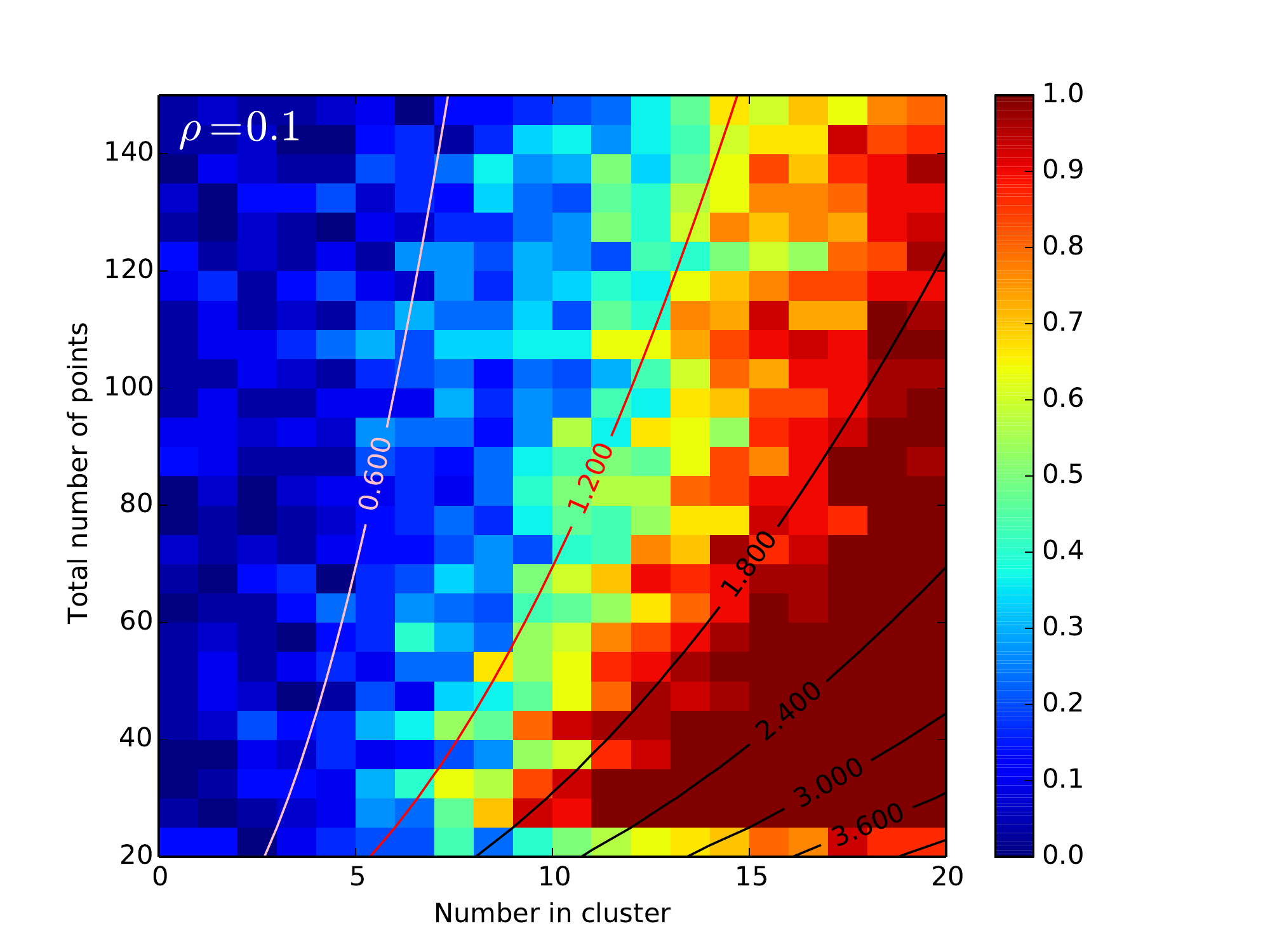}}
\subfloat{\includegraphics[width=0.5\linewidth]{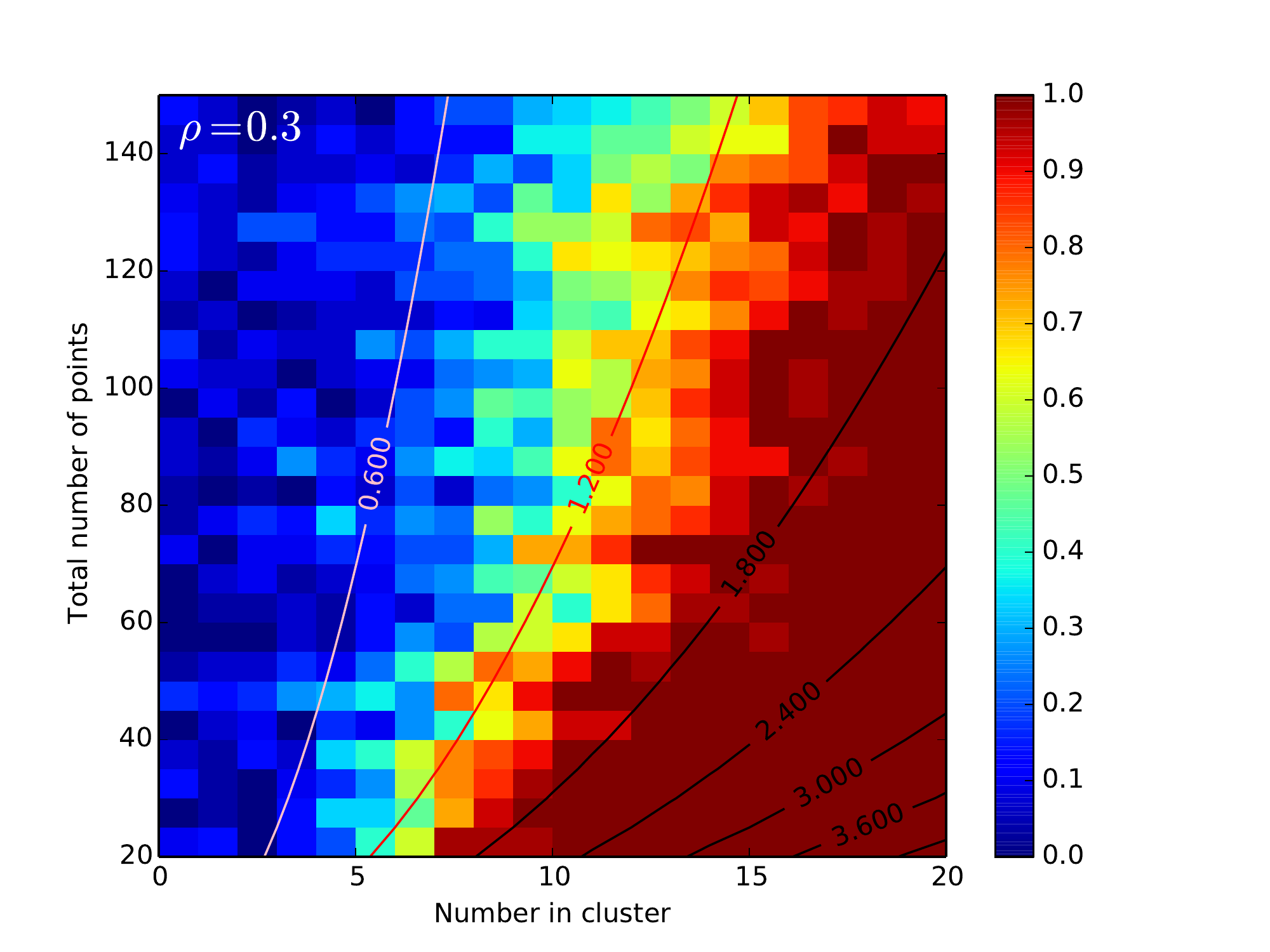}}\\
\subfloat{\includegraphics[width=0.5\linewidth]{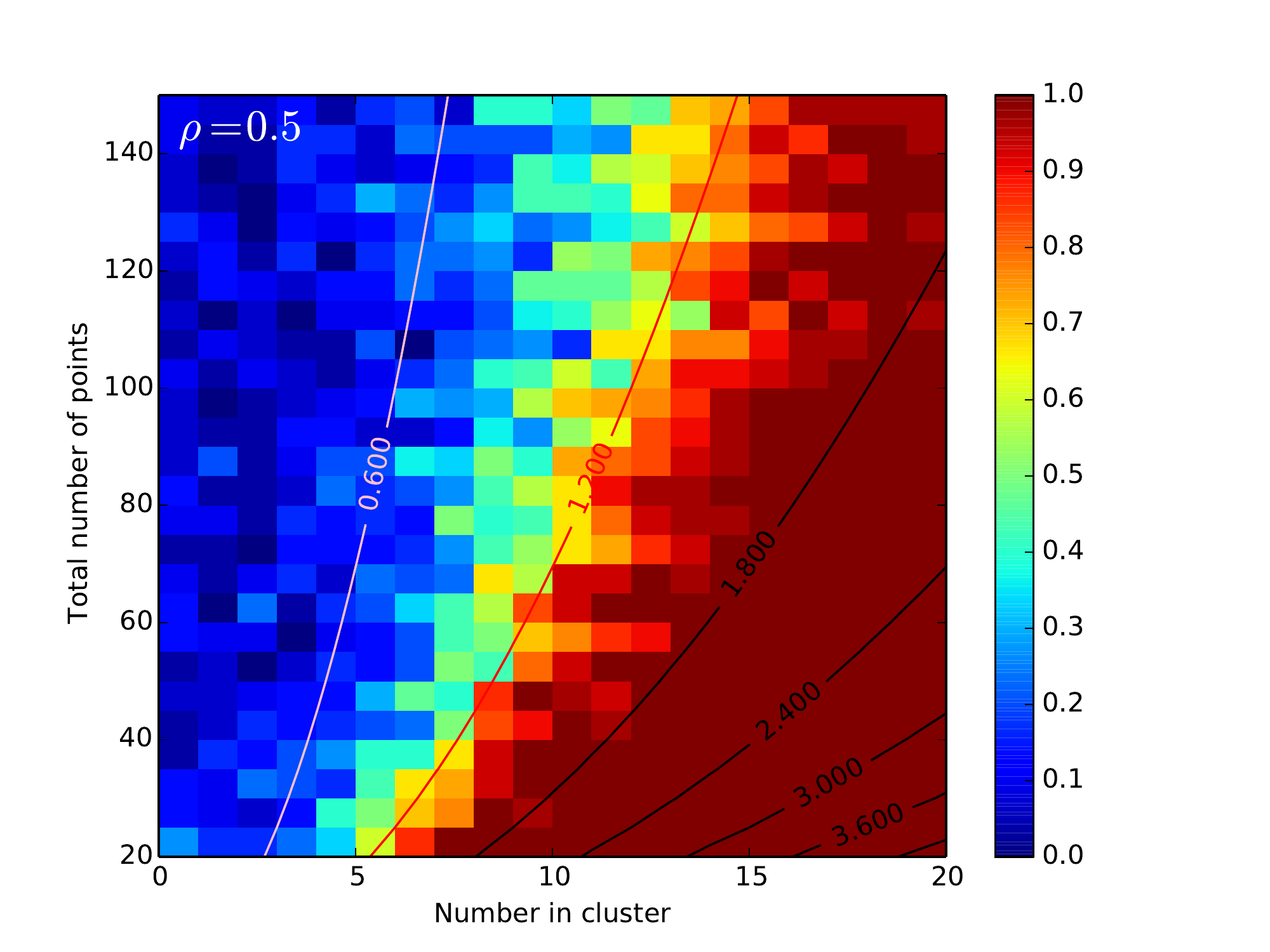}}
\subfloat{\includegraphics[width=0.5\linewidth]{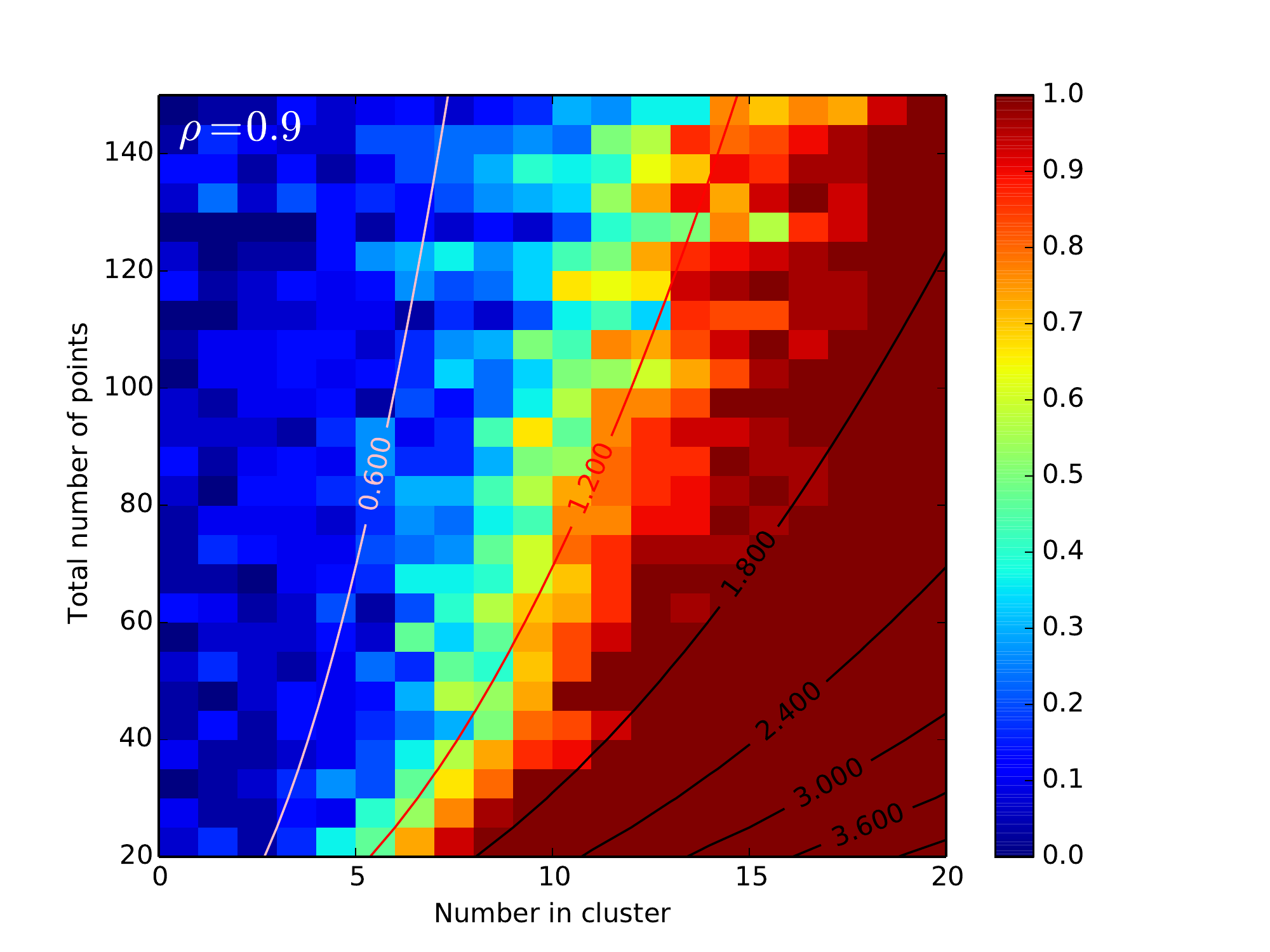}}
\caption{The rejection fraction with $P(H_0)<5\%$ for O-ring statistic with logarithmic annuli widths.  Within each panel, the annulus width is given by Equation \ref{eqn:logbin} and the contours show the theoretical SNR (Eqn. \ref{eqn:SNR}).  Between panels, $\rho$ increases from top left to bottom right with (a) $\rho=0.1$; (b) $\rho=0.3$, (c) $\rho=0.5$; and (d) $\rho=0.9$.}
\label{fig:logWidth}
\end{figure*}

An alternative method for the annuli width is to use logarithmic widths, where the ratio of the outer to inner radius of each annulus is kept constant, making the width a function of the radial distance. Fig. \ref{fig:logWidth} demonstrates four of these functions with half-widths given by 
\begin{equation}
\label{eqn:logbin}
q = \rho r
\end{equation} 
for $\rho = 0.1, 0.3, 0.5, \text{and } 0.9$. While there is an effect on the rejection fraction due to the bin width, as seen in Fig. \ref{fig:logWidth}, the results are consistent when $\rho \geq 0.3$. This is a promising result as this method requires no prior knowledge of the cluster width and no dependence on the number of points in the study region.

\subsection{Application to Astronomical Data}
\label{subsec: AstroResults}
\begin{figure*}
\centering
\subfloat{\includegraphics[width=0.5\linewidth]{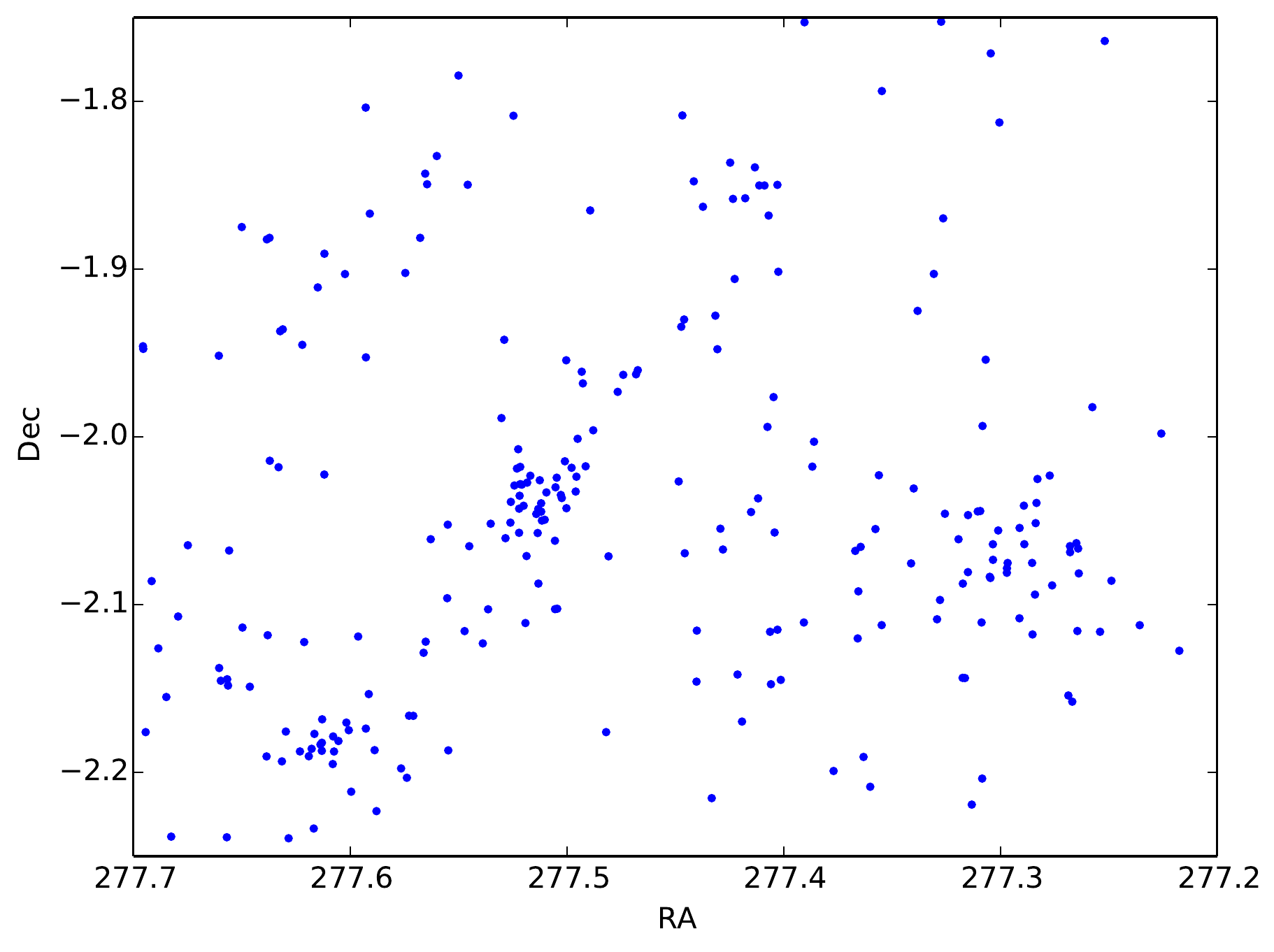}}\\
\subfloat{\includegraphics[width=0.5\linewidth]{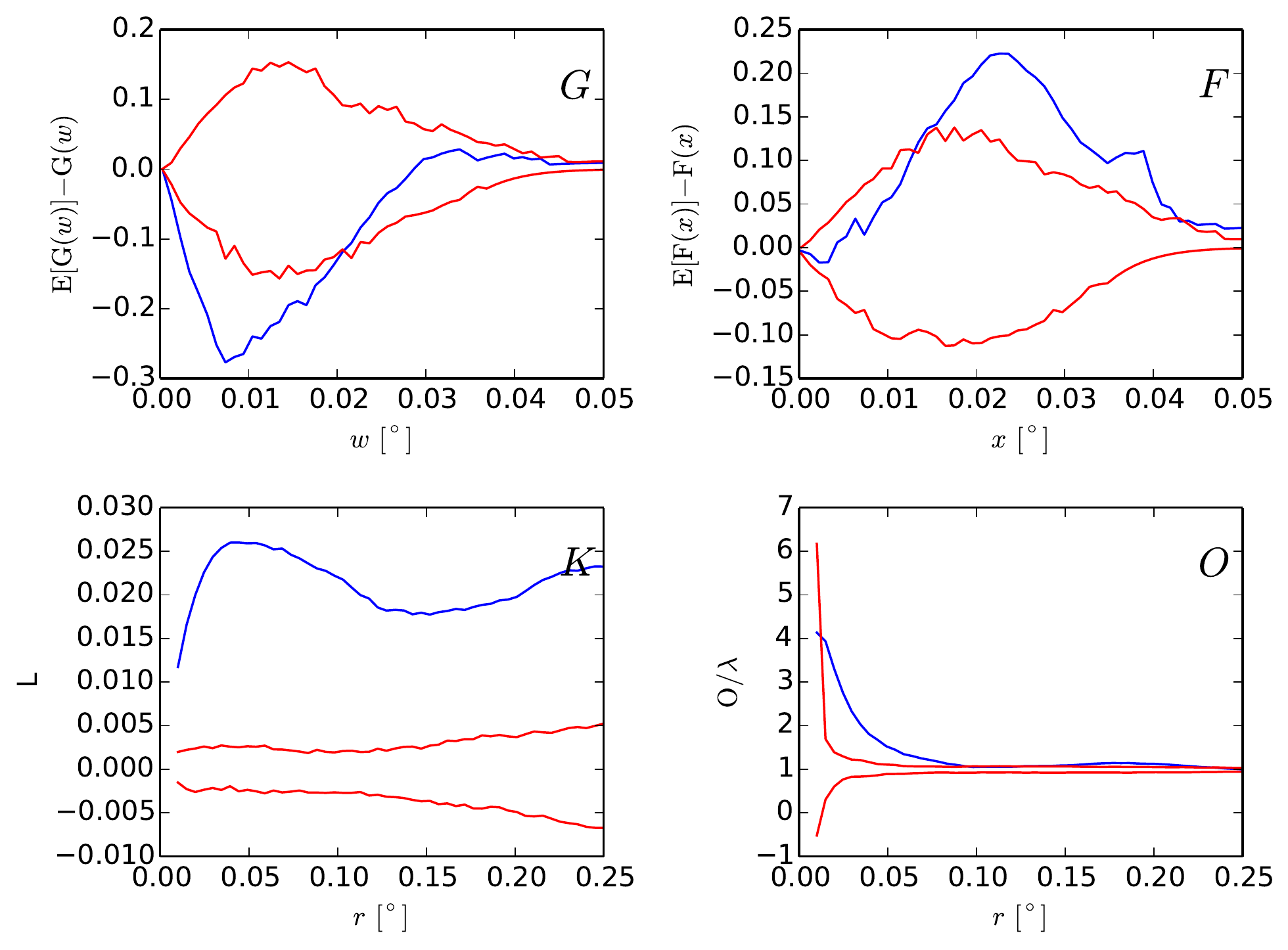}}
\caption{(above) Positions of YSOs within Serpens South, (below) results of G, F, K and O-ring with 95 per cent global confidence envelopes for CSR.}
\label{fig:serpens_yso_stats}
\end{figure*}

In addition to the simulated data tested above, three sets of astrophysical data were tested. The first was the locations of Young Stellar Objects around the Serpens South star forming region \citep{Gutermuth2008} from the \citet{dunham2015} catalogue, using the limits $277.2 \leq \mathrm{RA} \leq 277.7$ and $-2.25 \leq \hbox{Dec} \leq -1.75$. Fig. \ref{fig:serpens_yso_stats} shows the locations of the 246 YSOs and the results of applying G, F, K and O-ring. Each summary statistic exceeds the 95 per cent confidence envelope, therefore rejecting CSR as an appropriate model for the distribution of the YSOs with 95 per cent confidence.

\begin{br2}The second test on real data was performed on 2601 randomly chosen members of the \textit{Spitzer} catalogue
\footnote{The full \textit{Spitzer} Gould Belt Survey catalogue of infrared sources produced using the \textit{Spitzer} Cores to Disks (c2d) methodology: see  Harvey et al. 2007; Evans et al. 2007 available from \url{https://irsa.ipac.caltech.edu/data/SPITZER/C2D/doc/c2d_del_document.ps}}
 from the same study window. This was the number of \textit{Spitzer} Sources identified with the object type `star\textunderscore F0I' and, as Ripley's K is invariant to random thinning, allowed for a less computationally intensive method of testing the region as a whole. Using a laptop with a 2.5 GHz Intel Core i5 processor and 8 GB of 1600 MHz DDR3 memory the code, written in Python, calculated G, F, Ripley's K and O-ring statistics for 2601 objects tested at 100 radial scales in approximately 209 seconds using a single core. \end{br2} 
 
\begin{figure*}
\subfloat{\includegraphics[width=0.5\linewidth]{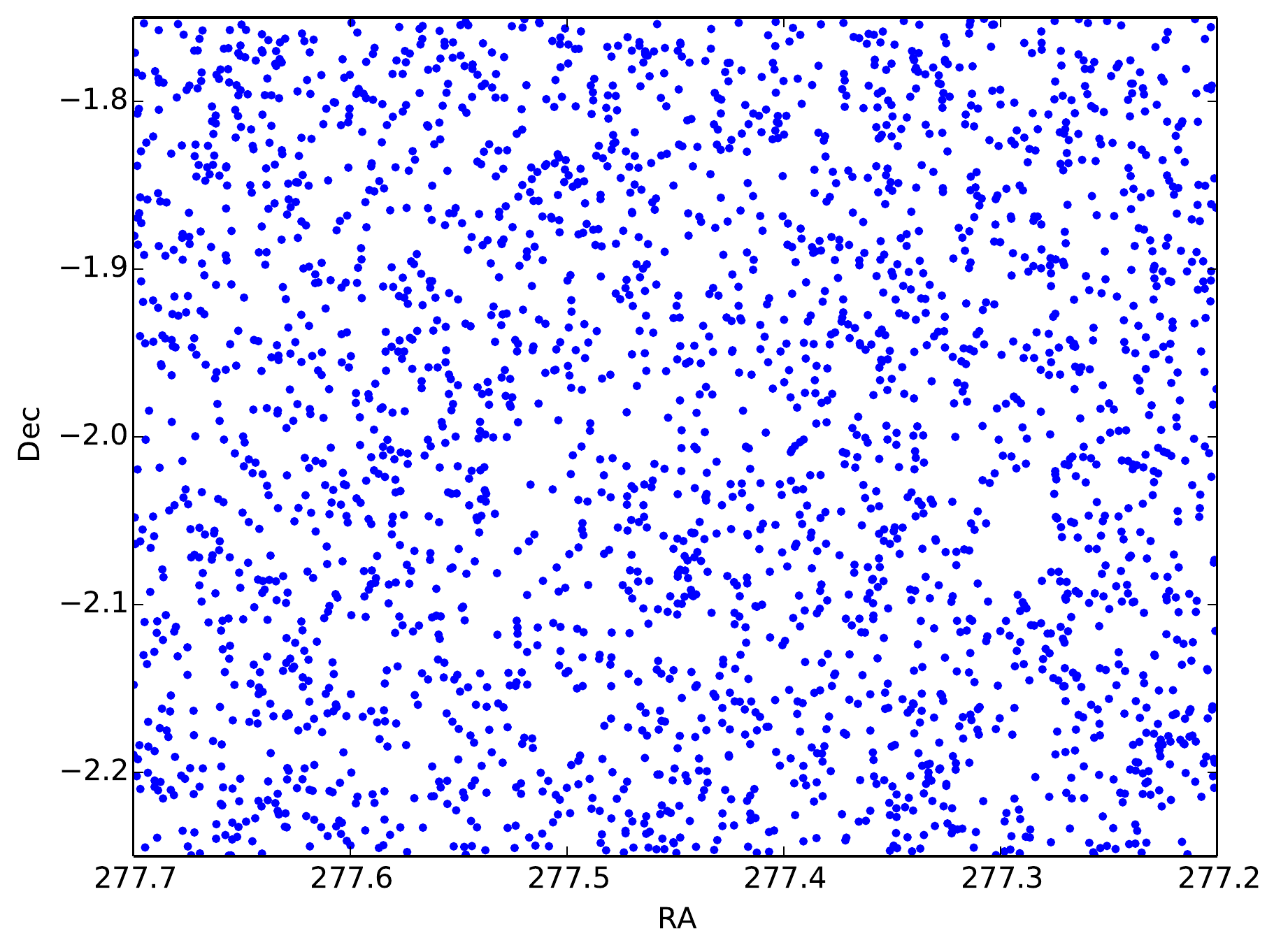}}
\subfloat{\includegraphics[width=0.5\linewidth]{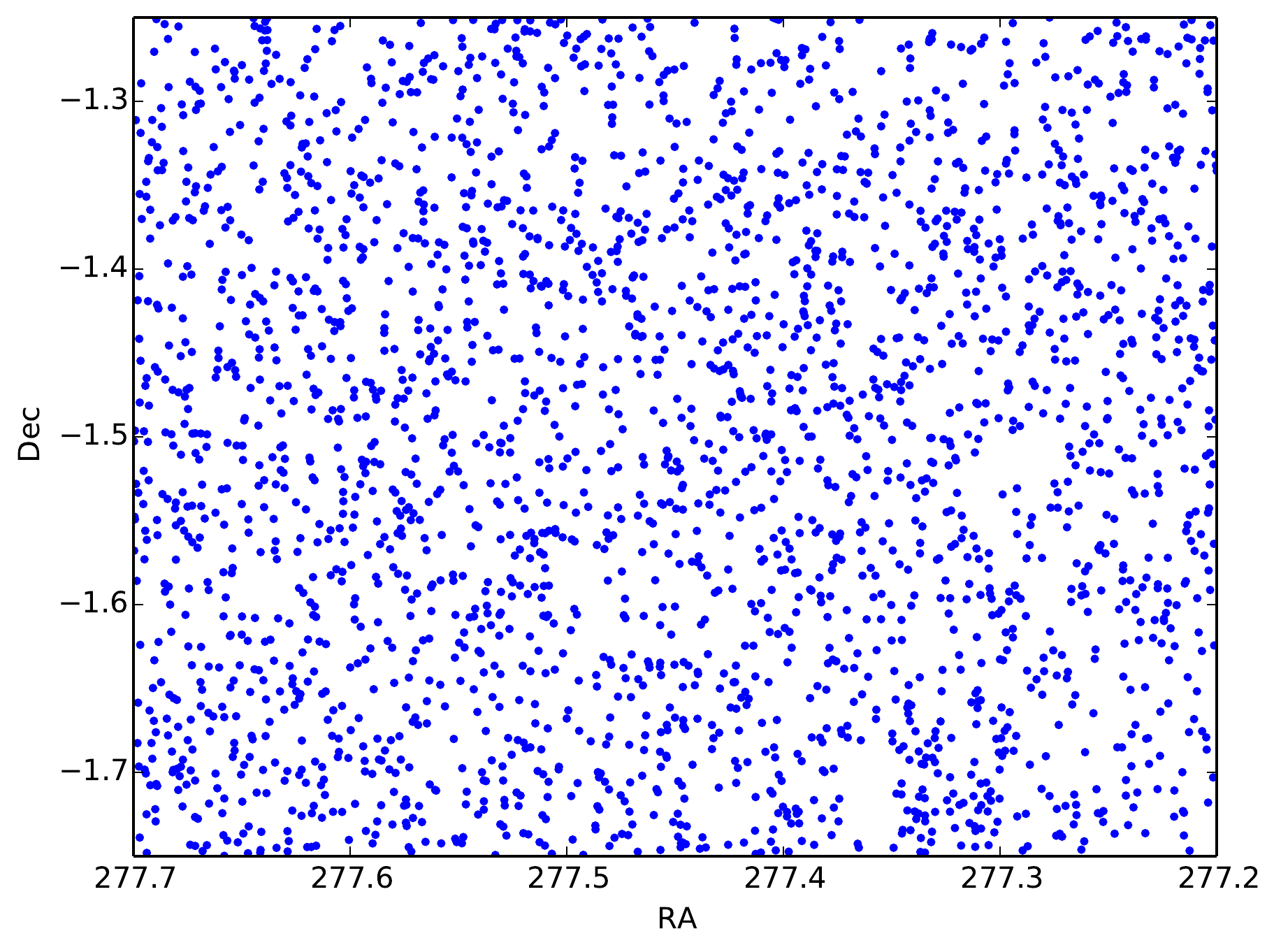}}\\
\subfloat{\includegraphics[width=0.5\linewidth]{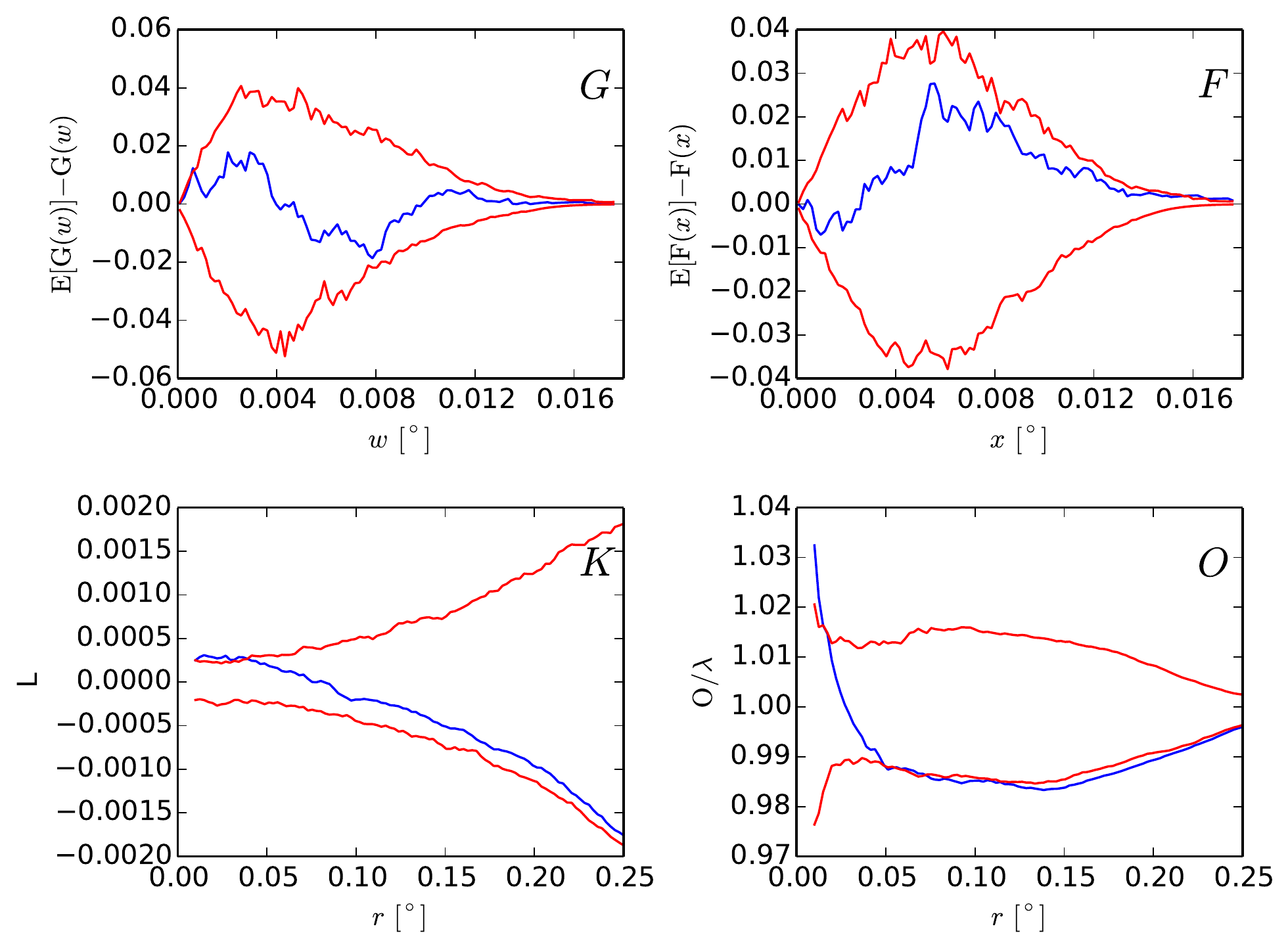}}
\subfloat{\includegraphics[width=0.5\linewidth]{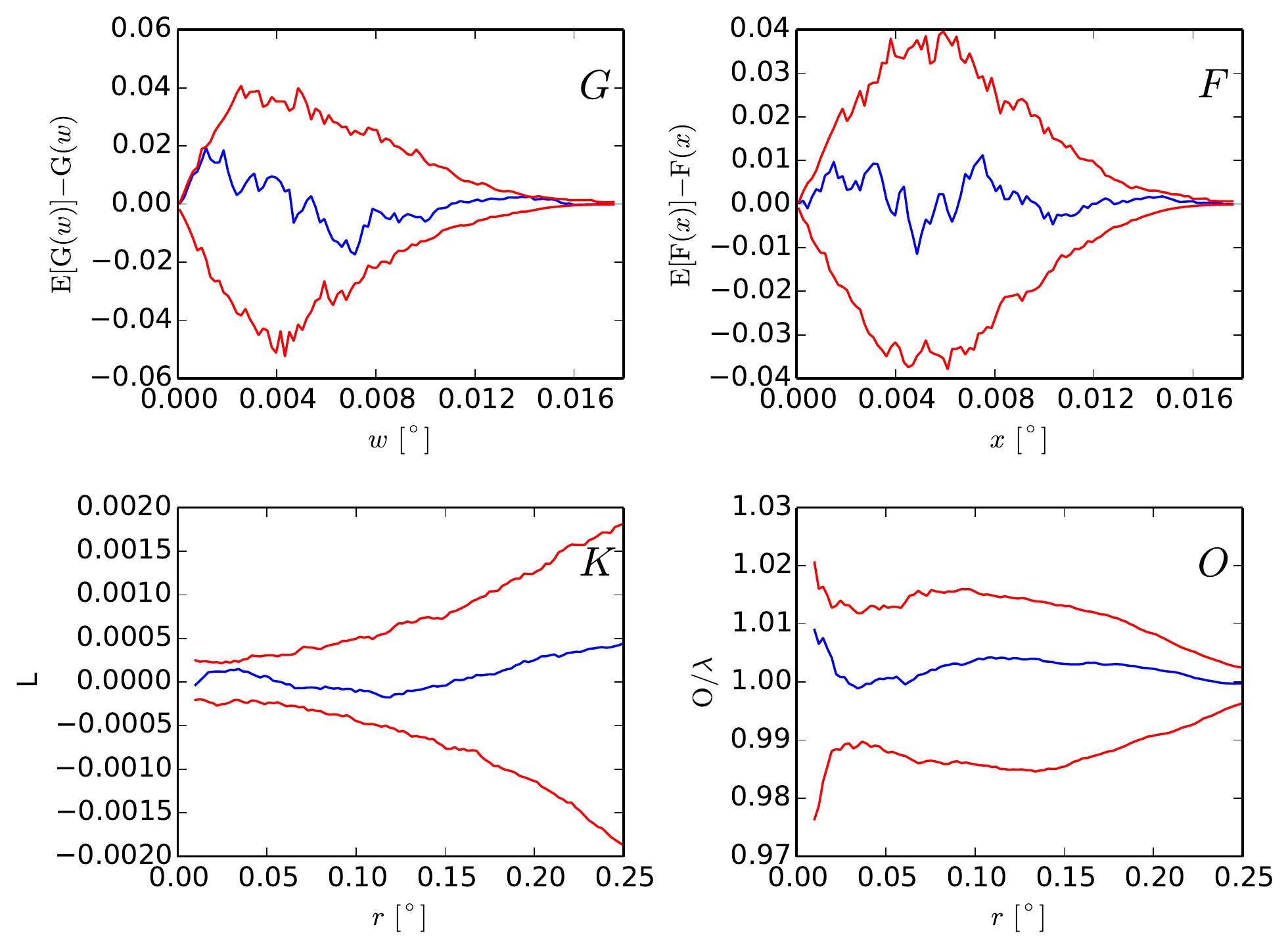}}
\caption{Left: (above) Positions of 2601 \textit{Spitzer} catalogue members within Serpens South; (below) results of G, F, K and O-ring with 95 per cent global confidence envelopes for CSR. Right: (above) Positions of 2601 \textit{Spitzer} catalogue members offset from Serpens South by $0.5^{\circ}$ in Dec ($277.2 \leq \mathrm{RA} \leq 277.7$ and $-1.75 \leq \hbox{Dec} \leq -1.25$); (below) G, F, K and O-ring results with 95 per cent global confidence envelope for CSR.}
\label{fig:serpens_spitzer_stats}
\end{figure*}

\begin{br}Fig. \ref{fig:serpens_spitzer_stats} (left) shows the positions of the randomly chosen members as well as the results from G, F K and O-ring. K and O-ring both reject CSR with 95 per cent confidence and display clustering followed by inhibition.\end{br} It is likely that this is due to the extinction of the cloud, the outline of which can be faintly seen in Fig. \ref{fig:serpens_spitzer_stats} (left) when compared to Fig. \ref{fig:serpens_yso_stats}. \begin{br}The third test consisted of 2601 random members chosen from an off-cloud region\end{br} translated in declination by $0.5^{\circ}$, shown in Fig. \ref{fig:serpens_spitzer_stats} (right). The results show that CSR cannot be rejected as a null hypothesis for the distribution of these randomly chosen members.

\begin{table*}
\caption{\label{table:decto_scores}Detectability scores for all tests.}
\begin{center}
\begin{tabular}{c c c c }
\hline
Cluster Radius & Test & Variation & Score\\
\hline
3 & G & & $0.174 \pm 0.004$\\
" & F & & $0.089 \pm 0.001$ \\
" & K & & $0.550 \pm 0.003$ \\
" & O-ring & variable, $\rho = 0.1$ & 0.53 \\
" & O-ring & variable, $\rho = 0.2$ & 0.56 \\
" & O-ring & fixed, $\rho = 0.1$ & 0.54 \\
" & O-ring & fixed, $\rho = 0.3$ & 0.59 \\
" & O-ring & fixed, $\rho = 0.5$ & 0.59 \\
" & O-ring & fixed, $\rho = 1.0$ & 0.54 \\
" & O-ring & logarithmic, $\rho = 0.1$ & 0.49 \\
" & O-ring & logarithmic, $\rho = 0.3$ & 0.56 \\
" & O-ring & logarithmic, $\rho = 0.5$ & 0.58 \\
" & O-ring & logarithmic, $\rho = 0.9$ & 0.56 \\
" & MST & $P(H_0)<5\%$ & $0.239  \pm 0.001$ \\
5 & G & & $0.106 \pm 0.006$ \\
" & F & & $0.069 \pm 0.003$ \\
" & K & & $0.438 \pm 0.003$ \\
" & O-ring & fixed, $\rho = 0.3$ & 0.48 \\
" & O-ring & fixed, $\rho = 0.5$ & 0.50 \\
" & MST & $P(H_0)<5\%$ & $0.179 \pm 0.001$\\
\hline
\end{tabular}
\end{center}
\end{table*}

\section{DISCUSSION}
\label{sec:discussion}
\subsection{Comparison between methods}
All four tests are capable of determining if an underlying process is random through significance testing. While additional tests are required to determine what type of clustering is present, a rejection of CSR due to a higher-than-expected average density is sufficient to determine the presence of overdensities within the data set, as well as an indication of the spatial scales. From the results in Section \ref{sec:Results}, the tests which reject randomness for this scenario the most sensitively are the second-order tests Ripley's K and O-ring, followed by G and then F. 

G is able to reliably reject CSR when the majority of the points belong to the cluster, but the likelihood of rejection drops off rapidly with increasing number of background stars. \begin{br2}\begin{br}F, however, shows few detections overall and is therefore an inappropriate test for CSR in this scenario.\end{br}\end{br2} G and F test the distribution of observed first nearest neighbour distances to distributions from realisations of CSR and to reject randomness there must be a significant shift in the distribution. Clustering produces a shift towards shorter nearest neighbour lengths, therefore the less-clustered and \begin{br}smaller-population clusters have a weaker effect on the distribution and are therefore more difficult to detect.\end{br} G has an advantage in this test as the test positions are the stellar positions, while F utilises random positions in the window which are not necessarily located close to, or inside, the cluster. 

The results of Ripley's K and O-ring are comparable, with the main difference being O-ring's dependence on bin width. As demonstrated by the detectability scores Ripley's K is the most consistent with an average score of $0.550 \pm 0.003$ while O-ring has a greater potential of CSR rejection when the bin width is optimised. With regards to choosing the bin width the results show that the logarithmic bins outperform Ripley's K and the variable bin widths for values of $\rho \geq 0.3$, and neither are as effective as a bin width matched to the cluster radius. For situations where a characteristic scale for the cluster is known a width approximately equal to this scale is preferred. 

From the four tests discussed, the best test for CSR rejection is Ripley's K due to its lack of dependence on any other parameters. However, the results from Ripley's K are less easy to interpret as described in Section \ref{subsec:envelope}.
Therefore, because O-ring is able to match or exceed this performance with most logarithmic bin widths, and because the results describe the density at and around a given radial distance, O-ring is recommended as the preferred test for this situation. 

Both of the second-order tests show a characteristic region of consistent rejection of CSR followed by a gradient towards non-rejection. The rejection fraction trends in a similar fashion to the contours of constant signal-to-noise given by Equation \ref{eqn:SNR} even when the overall detectability decreases due to having a more dispersed cluster, as seen in Fig. \ref{fig:compare3n5}. This is indicative of the way the tests function. By using all of the interpoint distances the effect on the density up to or at a given distance due to a cluster will be lessened by having a larger cluster which is more dispersed. Similarly, by increasing the size of the study region to include more background stars \begin{br2}the effect of a cluster on the measurements of K and O-ring will be reduced, and the statistics will be averaged closer to the background value, making it more difficult to exceed the envelope.\end{br2}

\subsection{Comparison to other tests} 
Two of the most commonly used methods to determine clustering or randomness in astrophysics are the two-point correlation function (2PCF) and the minimum spanning tree (MST).

\subsubsection{Two-Point Correlation Function}
The O-ring test (also known as MSDC) is related to the 2PCF $\xi (r)$ by \citep{Cressie1993}
\begin{equation}
\mathrm{O}(r) = (1+\xi (r))\lambda.
\end{equation}
The tests are therefore equivalent and it is the usage of the test that is the main difference. The results of the 2PCF are generally used to ascertain a radial dependence on density, and by inspecting this radial dependence describe the underlying mechanism that could have produced the results. In contrast the methods used in this paper simulate an underlying mechanism and determine how well the results of the mechanism fit the data. For example, the MSDC for the Taurus, Ophiucus and Orion Trapezium star forming regions shows two distinct clustering regimes \citep{Simon1997}, one for small-scale clustering of binaries and another for larger scale clustering, between which is a break point that varies for each star forming region. This is consistent with an underlying fractal structure over a subset of stellar densities, however, as \citet{bateclark1998} demonstrates, different distributing processes can produce MSDC lines that fit the data equally well.  

\subsubsection{Minimum Spanning Tree}
One method of using the MST to test for mass segregation by \citet{allison2009} has been adapted to test if a distribution of stars is random \citep{cantat-gaudin2018}. The \citet{cantat-gaudin2018} method is a one-sided test with significance level $\alpha = 0.16$ and so is not directly comparable with the other spatial statistics tests that have been performed. However, it was adapted into the following test such that the results may be compared with the other tests in the paper. 

\begin{figure}
\includegraphics[width=\columnwidth]{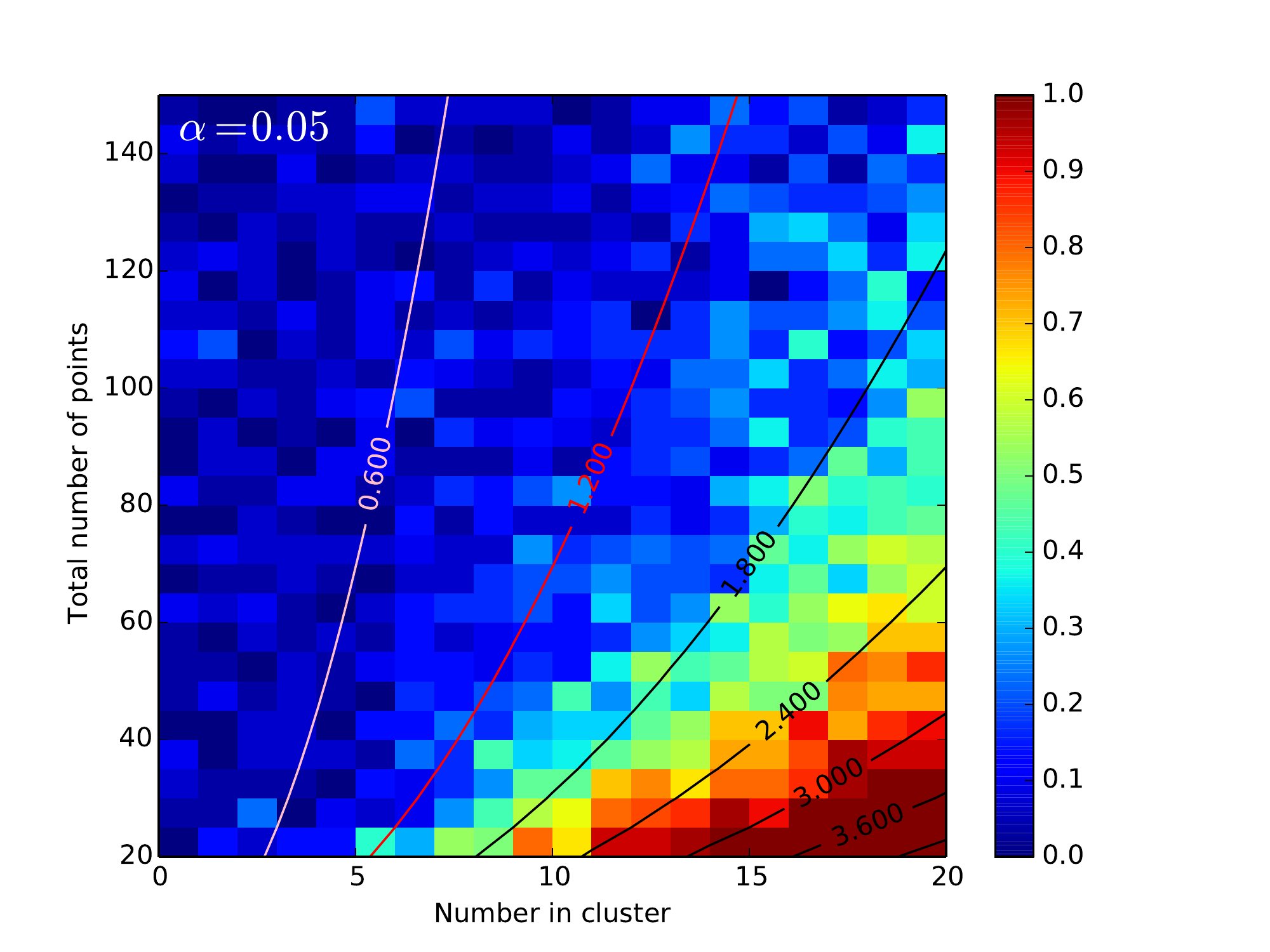}
\caption{The rejection fraction with $P(H_0)<5\%$ for MST total branch length test. The contours show the theoretical SNR (Eqn. \ref{eqn:SNR}).}
\label{fig:mst_results}
\end{figure}

The total branch length of $n$ realisations of CSR was recorded and ordered by size. Using the kth most extreme total branch lengths, as described in Section \ref{sec:significance}, produces a two-sided test with significance $\alpha = 2k/(n+1)$. For $n = 1999$ and $\alpha = 0.05$ the results are shown in Fig. \ref{fig:mst_results} and the detectability scores shown in Table \ref{table:decto_scores}. 

The detectability score of the test is comparable to Diggle's G function. This is to be expected as the minimum spanning tree contains all of the star-star nearest neighbour edges and G uses only the star-star nearest neighbour distances. This is one potential reason for it's performance being weaker than the second-order tests which utilise all $n(n-1)$ interpoint distances.

While it is less sensitive than the second-order tests, the MST does have some advantages which make it worth investigating further. As mentioned earlier there exist hierarchical clustering techniques to generate clusters \citep{Yu2015} which can identify cluster members which is something G, F, K and O-ring cannot do in the form described in this paper. Further, there are measures of the MST that exist such as the graph distance matrix, a symmetric matrix containing the number of edges on the path between any pair of points in the tree, which to our knowledge have not yet been applied to significance testing but could be useful in cluster identification and significance testing.


\subsection{Application to Astronomical Data}
In Section \ref{subsec: AstroResults} the distribution of YSOs within Serpens South and the on-cloud random sample of \textit{Spitzer} sources were shown to be inconsistent with CSR, while the random sample of off-cloud sources were shown to be unable to reject CSR as a null model for their distribution. The rejection of CSR runs for the YSO distribution was to be expected and demonstrates that the tests are able to reject genuine patterns that were not produced by randomly distributing stars in a window. It is interesting to note that O-ring shows the presence of two subsets of radial distances which exhibit overdensities though the effect causing this, either first or second-order, is not known as the distribution is not homogeneous and isotropic, so the appearance of clustering could be due to effects such as virtual aggregation as discussed in Section \ref{sec:Methods}. The rejection of CSR for the random on-cloud members due to extinction shows that these tests function not only for clustering processes but for inhibiting processes, while the off-cloud random members demonstrates that there are examples of astrophysical data which are consistent with CSR.

\section{CONCLUSIONS}
We have adapted spatial statistics methods for testing distributions of points from ecology and tested them for their ability to reject randomness. Centralised clusters were generated and projected on top of a population of randomly distributed background members and, by varying the number of stars in the cluster and background, a parameter space of the empirical probability of CSR rejection was produced for each statistical test. 

The best performing test was the O-ring test using logarithmically spaced overlapping bins. Ripley's K and the O-ring test with a constant bin width show a slightly reduced detectability compared to O-ring with logarithmic bins, making them both good tests but not optimal.

The O-ring test is equivalent to the two-point correlation function. Ripley's K and O-ring are more sensitive tests for CSR than the total branch length of the minimum spanning tree. 

The rejection of randomness for a given cluster radius approximately follows contours of signal-to-noise calculated over the parameter space of the number of stars in the cluster and background. A larger cluster radius over the same region of parameter space shows a decrease in the likelihood of rejection. 

Three example star fields were tested against complete spatial randomness. The null hypothesis of randomness was rejected for the distribution of young stellar objects within Serpens South, as well as randomly selected \textit{Spitzer} sources from the same region. The statistics from randomly selected \textit{Spitzer} sources sampled from an off-cloud region were consistent with complete spatial randomness.

\section*{ACKNOWLEDGEMENTS}
Brendan Retter is funded by an STFC studentship. 
This research has made use of the Starlink software \citep{2014ASPC..485..391C} which is supported by the East Asian Observatory. 
The figures in this paper have been produced using matplotlib: a 2D graphics package used for Python for application development, interactive scripting, and publication-quality image generation across user interfaces and operating systems \citep{Hunter:2007}. 
This research has made use of NASA's Astrophysics Data System.
This work is based (in part) on observations made with the Spitzer Space Telescope, which is operated by the Jet Propulsion Laboratory, California Institute of Technology under a contract with NASA. 

\begin{table}
\caption{Table of Symbols}
\label{table:symbols}
\begin{tabular}{ll}
\hline
Symbol & Description\\
\hline
$\alpha$ & significance level \\
$n$ & number of simulated patterns \\
$\lambda$ & first-order intensity \\
$\s$ & arbitrary areas in study region\\
$w$ & star-star nearest neighbour distance\\
$x$ & position-star nearest neighbour distance\\
$r$ & radial distances\\
$\A$ & study region\\
$q$ & half-width of annulus\\
$\rho$ & multiplicative factor for annulus width\\
$R$ & cluster radius\\
$N$ & number of points\\
$\Nc$ & number of points in cluster\\
$\Nbg$ & number of background points\\
$\Ntot$ & total number of points\\
$H_0$ & null hypothesis\\
\hline
\end{tabular}
\end{table}




\bibliographystyle{mnras}
\bibliography{YSOreading} 


\bsp	
\label{lastpage}
\end{document}